%% file: main.tex
\PassOptionsToPackage{table}{xcolor}

\documentclass[acmsmall]{acmart}
\AtBeginDocument{
  }

\makeatletter
\def\@ACM@copyright@check@cc{}
\makeatother
\setcopyright{cc}
\setcctype{by}
\acmJournal{POMACS}
\acmYear{2026} \acmVolume{10} \acmNumber{2} \acmArticle{37}
\acmMonth{6} \acmDOI{10.1145/3805635}

\newcommand{\myparagraph}[1]{\noindent{\bf {#1}.}}
\newcommand*\circled[1]{\tikz[baseline=(char.base)]{
            \node[shape=circle,draw,inner sep=1pt] (char) {#1};}}
\newcommand{\takeaway}[1]{
    \vspace{2mm}
    \noindent\fbox{\parbox{\columnwidth}{#1}
    }
}
\newcommand{\projectname}{ECCentric}

\usepackage{tikz}
\usepackage{hyperref}
\usepackage{quantikz}
\usepackage{makecell}
\graphicspath{ {./figures/} }
\usetikzlibrary{
    shapes,
    arrows,
    arrows.meta,
    decorations.pathreplacing,
    calc,
}
\usepackage{subcaption}
\usepackage{multirow}
\usepackage{float}
\usepackage{arydshln}
\usepackage{float}
\usepackage{pifont}
\usepackage{array}
\usepackage{xcolor}
\usepackage{wrapfig}
\usepackage{multirow}
\usepackage{hhline}
\usepackage{enumitem}
\usepackage{placeins}
\usepackage{nicematrix}
\newcommand{\cmark}{\textcolor{green!70!black}{\textbf{\checkmark}}}
\newcommand{\xmark}{\textcolor{red!80!black}{\textbf{\ding{55}}}}

\input{space-tricks}
\input{glossaries}

\begin{document}
\title{ECCentric: An Empirical Analysis of Quantum Error Correction Codes}

\author{Aleksandra Świerkowska}
\email{aleksandra.swierkowska@tum.de}
\orcid{0009-0008-8688-0155}
\affiliation{
 \institution{Technical University of Munich}
 \country{Germany}
}

\author{Jannik Pflieger}
\orcid{0009-0003-8186-4950}
\affiliation{
  \institution{Technical University of Munich}
  \country{Germany}
}

\author{Emmanouil Giortamis}
\orcid{0009-0000-3638-2969}
\affiliation{
  \institution{Technical University of Munich}
  \country{Germany}
}

\author{Pramod Bhatotia}
\orcid{0000-0002-3220-5735}
\affiliation{
  \institution{Technical University of Munich}
  \country{Germany}
}

\renewcommand{\shortauthors}{Aleksandra Świerkowska, Jannik Pflieger, Emmanouil Giortamis, and Pramod Bhatotia}

\input{chapters/abstract}



\received{January 2026}
\received[revised]{March 2026}
\received[accepted]{April 2026}

\maketitle

\glsresetall
\input{chapters/intro}

\input{chapters/background}
\input{chapters/overview}

\input{chapters/hardware_restriction_analysis}
\input{chapters/framework_analysis}

\input{chapters/noise_analysis}

\input{chapters/limitations}

\input{chapters/related_work}
\input{chapters/future_implications}

\begin{acks}    
We thank Dr. Joschka Roffe for his insights on the taxonomy of QEC codes and Peter Wegmann for his assistance with the implementation. Funded by the Bavarian State Ministry of Science and the Arts as part of the Munich Quantum Valley (MQV), grant number 6090181.
\end{acks}

\bibliographystyle{ACM-Reference-Format}
\bibliography{bibliography}

\end{document}

%% file: space-tricks.tex
\setlist{noitemsep,topsep=0pt,parsep=0pt,partopsep=0pt}

\usepackage[subtle]{savetrees}

\usepackage{setspace}
\usepackage[margin=5pt,font={stretch=0.9}]{caption}

%% file: glossaries.tex
\usepackage[acronym]{glossaries}
\makenoidxglossaries
\newacronym{bposd}{BP-OSD}{belief propagation with order statistic decoding}
\newacronym{dqc}{DQC}{Distributed Quantum Computing}
\newacronym{ft}{FT}{fault-tolerant}
\newacronym{ldpc}{LDPC}{Low-Density Parity-Check}
\newacronym{ml}{ML}{Machine Learning}
\newacronym{mwpm}{MWPM}{minimum-weight perfect matching}
\newacronym{qec}{QEC}{Quantum Error Correction}
\newacronym{qldpc}{QLDPC}{Quantum Low-Density Parity-Check}

\glsaddall

%% file: chapters/abstract.tex
\begin{abstract}

Quantum Error Correction (QEC) is essential for building scalable quantum computers, but a lack of systematic, end-to-end evaluation methods makes it difficult to assess how different QEC codes perform under realistic conditions. The vast diversity of codes, an expansive experimental search space, and the absence of a standardized framework prevent a thorough, holistic analysis. To address this, we introduce ECCentric, an end-to-end benchmarking framework designed to systematically evaluate QEC codes across the full quantum computing stack. ECCentric is designed to be modular, extensible, and general, allowing for a comprehensive analysis of QEC code families under varying hardware topologies, noise models, and compilation strategies.

Using ECCentric, we conduct the first systematic benchmarking of major QEC code families against realistic, mid-term quantum device parameters. Our empirical analysis reveals that intra-QPU execution significantly outperforms distributed methods, that qubit connectivity is a far more critical factor for reducing logical errors than increasing code distance, and that compiler overhead remains a major source of error. Furthermore, our findings suggest that trapped-ion architectures with qubit shuttling are the most promising near-term platforms and that on noisy devices, a strategic and selective application of QEC is necessary to avoid introducing more errors than are corrected. This study provides crucial, actionable insights for both hardware designers and practitioners, guiding the development of fault-tolerant quantum systems.

\end{abstract}

%% file: chapters/intro.tex
\section{Introduction}
\glsresetall

\myparagraph{Context and motivation}
While the promise of quantum computing \cite{Shor_factorization, grover1996fastquantummechanicalalgorithm, Preskill_2018, Arute_2019} has led to the rapid development and accessibility of Quantum Processing Units (QPUs) \cite{ibmQuantum, aws-quantum, google-quantum, azure-quantum, IBM_roadmap, Quantinuum2024Roadmap}, achieving true quantum advantage is fundamentally limited by persistent hardware noise \cite{Preskill_2018}.

Quantum Error Correction (QEC) offers the only known path to fault tolerance, a prerequisite for building scalable quantum computers capable of solving classically intractable problems \cite{Fowler_2012, Gidney_2021, aharonov1999faulttolerantquantumcomputationconstant}. It encodes a single logical qubit into a redundant, entangled system of multiple noisy physical qubits \cite{Preskill_2018, Nielsen_Chuang_2010}. Once such an encoded circuit is transpiled into the hardware’s native gate set and adapted to its topology (Fig.~\ref{fig:qec_pipeline_example}), it is executed on the device, where syndrome measurements are continuously performed, allowing the decoder to correct errors in real time \cite{Battistel_2023}.

Notably, a diverse family of QEC codes exists, each with unique resource overheads, operational assumptions, and performance characteristics. Their suitability heavily depends on the underlying physical qubit technology and its noise model~\cite{Fowler_2012, PRXQuantum.2.040101}.

\myparagraph{Research gap and challenges} Despite the variety of QEC codes, there's a significant research gap: a lack of a systematic, end-to-end methodology for rigorously comparing them. This makes it challenging to accurately assess a code's performance under the specific, realistic noise and software constraints of a particular QPU. A comprehensive practical evaluation needs to consider the entire execution pipeline, as each stage can influence the code's effectiveness.

\input{figures_tex/rq_spidergraph}

Unfortunately, current QEC evaluations are often ad hoc and limited, either focusing on a small number of codes with oversimplified noise models \cite{small_codes_on_hardware_2022, comparison_shor_vs_steane_on_beacon_trappedion_2024, comparison_11codes_2009, comparison_many_codes_huang_2019} or analyzing components such as decoders in isolation \cite{bechmarking_controller_kurman_2024, benchmarking_ml_qec_zhao_2024, heavyhex_2025, bacon_surface_shor_trapped_ion_2020, surface_vs_realistic_noise_2014, qpandora_chatterjee_2025, steane_noise_comparison_2015, comparison_hypergraph_qldpc_2020}. As a result, no existing benchmarking framework provides an end-to-end, systematic evaluation.
Current tools exhibit a distinct dichotomy: fast, specialized simulators, such as Stim \cite{framework_stim_2021}, are hardware-agnostic and rely on simplistic error models, whereas general-purpose frameworks, such as Qiskit \cite{javadiabhari2024quantumcomputingqiskit}, lack the scalability needed to simulate large, fault-tolerant codes \cite{Fowler_2012}. These limitations create a major blind spot in practical quantum computing.

To effectively address this research gap, we must confront three fundamental challenges. {\em Firstly, the diverse QEC landscape:} A vast number of codes exist, each with unique overheads, error thresholds, and decoding complexities, making it difficult to compare their effectiveness across different hardware and noise models \cite{survey_noise_adapted_2022, Bravyi2024, wang2025demonstrationlowoverheadquantumerror, magic_mirror_2024, bacon_surface_shor_trapped_ion_2020, ye2025quantumerrorcorrectionlong}. {\em Secondly, the vast experimental search space:} A code's performance is determined by a multi-dimensional interplay of QPU topology, physical noise, and compiler artifacts, rendering an exhaustive search computationally intractable \cite{Benito2025comparativestudyof, surface_vs_realistic_noise_2014, variational_compiler_2021, qecc_synth_2025}. {\em Finally, the lack of a suitable benchmarking framework:} Existing tools are either too slow for large-scale studies (e.g., Qiskit \cite{javadiabhari2024quantumcomputingqiskit}) or too hardware-agnostic for realistic evaluation (e.g., Stim \cite{framework_stim_2021}). A viable solution must be general, supporting a wide spectrum of code families and QPU architectures; extensible, to easily integrate new decoders and evolving noise models; and modular, to enable the isolation and analysis of specific stages within the QEC pipeline, from encoding to decoding.

\input{figures_tex/qec_pipeline_example}

\myparagraph{Research question and our approach}
These challenges lead to our central research question: {\em How can we systematically evaluate the suitability of quantum error correction (QEC) codes for practical applications on current and near-term quantum devices?}

To answer this question, we propose a two-staged approach. First, we impose structure on the diverse and rapidly evolving QEC landscape by creating a systematic taxonomy of QEC codes. From this taxonomy, we have identified five primary families of codes and selected a representative member from each to ensure our analysis is both fair and comprehensive.

Building on this foundational taxonomy, we introduce \projectname{}, a new, end-to-end benchmarking framework designed for the systematic evaluation of QEC codes on current and near-term quantum devices.
\projectname{} is general, capable of supporting a wide array of QEC codes, QPU technologies, and compiler back-ends. It is modular, allowing researchers to customize the evaluation pipeline and scope of their experiments. And it is extensible, making it easy to integrate new code, hardware features, and compilation strategies as they emerge.

Leveraging the full capabilities of the ECCentric framework, we structure our investigation around four key dimensions (presented in Fig.~\ref{fig:axes}) that influence QEC performance, defining two targeted research questions for each to systematically assess their impact. These four dimensions include {\em QEC codes}, where we select six representative codes from all major families using our QEC code taxonomy; {\em QPU noise models}, where we include both ideal and realistic models derived from current (e.g., Google Willow \cite{Willow2025}) and next-generation (e.g., Quantinuum Apollo \cite{Quantinuum2024Roadmap} and IBM Flamingo \cite{IBM_roadmap}) quantum devices; {\em QPU topologies}, where we explore a range of layouts, from abstract to the complex architectures of existing hardware (e.g., Google Willow \cite{Willow2025} and Inflection \cite{radnaev2025universalneutralatomquantumcomputer}), varying qubit count and connectivity to isolate the impact of topology; and finally, {\em quantum compilation}, where we investigate how different quantum compilers (e.g., Qiskit \cite{javadiabhari2024quantumcomputingqiskit} and TKET \cite{Sivarajah_2020_tket}) and their internal stages, such as mapping and routing heuristics (e.g., SABRE \cite{gushu2019sabre}), affect QEC performance.

\input{figures_tex/qec_pipeline}

\myparagraph{Our key findings, implications, and lessons learned}  Our systematic evaluation reveals several critical insights and implications for both hardware and software development:

\begin{itemize}[leftmargin=*]

\item {\em Fault-tolerance with Trapped-Ion is near:} Our analysis of projected hardware roadmaps shows that the planned error rates and qubit shuttling capabilities of these devices should eliminate logical errors across most evaluated codes within the next five years \cite{Quantinuum2024Roadmap}.

\item {\em Intra-QPU execution is superior:} Our results show that distributing codes across QPUs increases the logical error rate by {69.14\%} compared to single-QPU execution. However, while link fidelity is a factor, the principal issue is the low number of connections, and hardware development should prioritize increasing the cross-QPU links rather than solely focusing on their fidelity.

\item {\em Prioritize connectivity over code distance:} Our experiments show that increasing code distance is often ineffective, increasing the logical error rate on average by 0.012 with only 30.95\% of adjustments yielding meaningful improvement. In contrast, improving connectivity provides substantial gains: moving from a grid to a fully connected topology reduces the logical error rate by {81.9\%}, while devices with qubit shuttling outperform those without by {45.48\%}. Thus, hardware manufacturers should prioritize improving qubit connectivity.

\item {\em QEC-aware compilation is essential:} The compilation process introduces significant overhead that can undermine the benefits of error correction. We find that mapping and routing add an average of 134.095\% more two-qubit gates, while optimized translation still adds 3.166 extra gates per original gate. Mitigating this requires QEC-aware compilers that optimize at the logical-qubit level, for example, by prioritizing efficient routing to busy ancilla qubits and canceling redundant gate sequences across repeated QEC cycles.

\item {\em Heterogeneity is not a primary concern:} We find the variability in individual qubit quality or in readout correctness has a negligible impact on performance, changing the logical error rate by 0.02 on average. This suggests that complex, variance-aware compilation strategies might be an unnecessary burden. This frees compiler developers to de-prioritize these intricate techniques and instead pursue more straightforward and effective designs.

\item {\em Apply QEC strategically, not universally:} On noisy, near-term devices, the indiscriminate application of QEC can be counterproductive. Our results show that at a physical two-qubit error rate of 0.002, most of our evaluated codes fail, and at 0.004, none remain effective, introducing more errors than they correct. This necessitates a selective approach where QEC is only used for operations or qubits where its benefits outweigh its overhead.

\end{itemize}

\myparagraph{Our contributions}
Overall, to the best of our knowledge, this paper provides the first systematic benchmarking of multiple QEC codes, focusing on their practical application on real-world hardware, compilers, and noise models. Our contributions are summarized as follows:
\begin{itemize}[leftmargin=*]
 \item {\em   Experimental QEC code analysis:} A systematic benchmarking of the major \gls{qec} code families under realistic experimental conditions.
 \item {\em  \projectname{} framework:}  The design of a modular benchmarking framework that can be readily extended to incorporate additional codes, devices, noise models, and decoders.
 \item {\em   QEC code taxonomy:} The introduction of a structured taxonomy for \gls{qec} codes, clearly presenting the differences and shared characteristics among the primary families.
 \item {\em Concatenated Steane code extension:} The first extension of the $[[49, 1, 9]]$ concatenated Steane code \cite{Pato_2024} to $[[373, 1, 27]]$ concatenated Steane code.
\end{itemize}

%% file: figures_tex/rq_spidergraph.tex
\begin{wrapfigure}{r}{0.4\textwidth}
    \centering
    \resizebox{0.37\textwidth}{!}{
    \begin{tikzpicture}[scale=0.87, every node/.style={font=\footnotesize}]
        \tikzset{
            mainnode/.style={
                diamond, draw, fill=#1!15,
                minimum width=1.8cm,
                minimum height=1.8cm,
                inner sep=2pt,
                align=center,
                text width=1.8cm
            },
            rqn/.style={
                circle, draw, fill=#1!20,
                minimum size=1.0cm,
                inner sep=0pt,
                align=center,
                text width=1.6cm
            }
        }

        \begin{scope}[rotate=0]

            \node[circle,draw,fill=white!15,minimum size=0.001cm] (C) {};

            \node[mainnode=blue] (Codes) at (90:1.4cm) {\textbf{QEC codes}};
            \node[mainnode=green] (Noise) at (0:1.4cm) {\textbf{Noise models}};
            \node[mainnode=red] (Topo) at (270:1.4cm) {\textbf{Topologies}};
            \node[mainnode=yellow] (Comp) at (180:1.4cm) {\textbf{Compilation}};

            \foreach \n in {Codes,Noise,Topo,Comp}{
                \draw[thick] (C) -- (\n);
            }

            \node[rqn=blue] (RQ1) at (74:3.2cm) {Error suppression};
            \node[rqn=blue] (RQ2) at (106:3.2cm) {Resource needs};
            
            \node[rqn=green] (RQ3) at (344:3.2cm) {Correction overhead};
            \node[rqn=green] (RQ4) at (16:3.2cm) {Error distribution};
            
            \node[rqn=red] (RQ5) at (254:3.2cm) {Qubit count};
            \node[rqn=red] (RQ6) at (286:3.2cm) {Qubit connectivity};
            
            \node[rqn=yellow] (RQ7) at (164:3.2cm) {Gate translation};
            \node[rqn=yellow] (RQ8) at (196:3.2cm) {Circuit mapping};

            \draw (RQ1) -- (Codes);
            \draw (RQ2) -- (Codes);

            \draw (RQ3) -- (Noise);
            \draw (RQ4) -- (Noise);

            \draw (RQ5) -- (Topo);
            \draw (RQ6) -- (Topo);

            \draw (RQ7) -- (Comp);
            \draw (RQ8) -- (Comp);

        \end{scope}
    \end{tikzpicture}
    }
    \caption{Our comprehensive framework provides a structured empirical analysis of QEC landscape. We examine the QEC field across four key dimensions, addressing eight critical research questions.}
    
    \label{fig:axes}
\end{wrapfigure}

%% file: figures_tex/qec_pipeline_example.tex
\begin{figure}[t]
    \centering
    \includegraphics[scale=0.3]{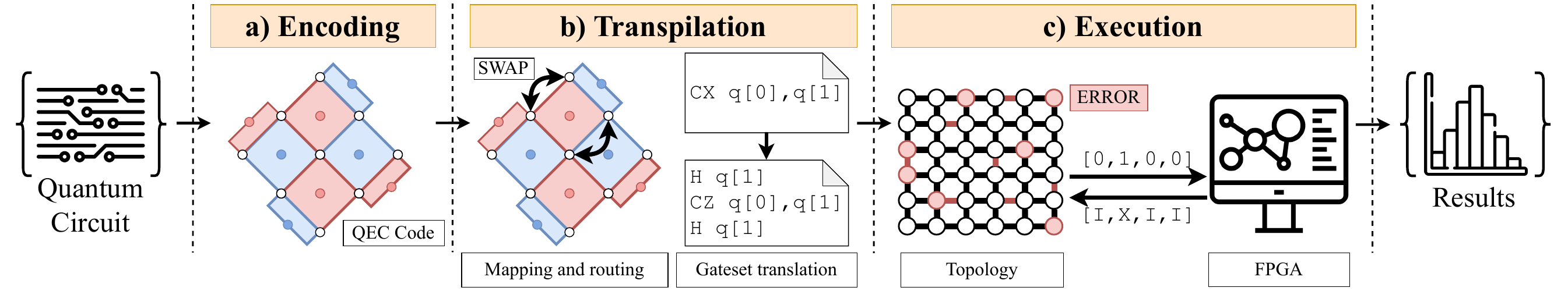}
    \caption{Workflow of a quantum circuit protected with QEC, showing how mapping and routing, translation, and decoding are handled during execution.}
    \label{fig:qec_pipeline_example}
\end{figure}

%% file: figures_tex/qec_pipeline.tex
\begin{figure}[t]
    \centering
    \includegraphics[scale=0.25]{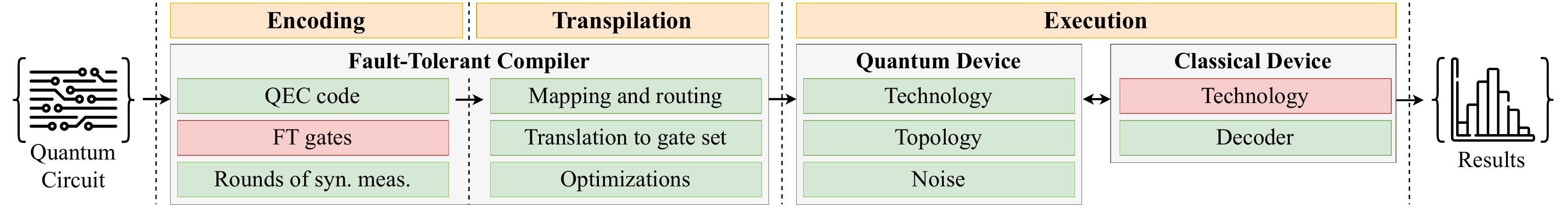}
    \caption{Lifetime of a quantum circuit protected with QEC code. {\em Our analysis covers the elements in green.
    }}
    \label{fig:theoretical_pipeline}
\end{figure}

%% file: chapters/background.tex
\section{Quantum Error Correction (QEC)}
\label{sec:background}
\input{tables/technologies}

QEC is a technique of protecting quantum information from noise by encoding it into entangled states of multiple qubits, i.e., logical qubits \cite{Nielsen_Chuang_2010}. Preparing and running such protected circuits, however, requires a complex toolchain that extends well beyond standard compilation. To the best of our knowledge, no fully integrated end-to-end compiler for QEC exists, but building on existing frameworks~\cite{Leblond_2023, watkins2024high}, we illustrate in Fig.~\ref{fig:theoretical_pipeline} the stages of compiling and executing a quantum circuit with QEC, each of which can introduce overhead and weaken protection:

\myparagraph{Encoding}
To provide protection for a quantum circuit (Fig.~\ref{fig:qec_pipeline_example}(a)), we start by choosing a QEC code matching the constraints of the device regarding topology and noise profile, and preparing logical qubits by entangling them with themselves and with ancilla qubits accordingly. Next, we translate the gates in the circuit into fault-tolerant (FT) gates able to manipulate logical qubits without falsely marking errors \cite{Nielsen_Chuang_2010}. While some gates can be translated with relatively low overhead, they alone cannot form a universal set \cite{Eastin_2009}. Missing gates need to be added with costly techniques such as code switching \cite{code_switching}, lattice surgery \cite{lattice_surgery}, or magic state distillation \cite{magic_state_destillation}. At present, no general tool can automatically translate an arbitrary circuit into a fully encoded form with support for these gates. Afterwards, we insert rounds of syndrome measurement periodically into the circuit, i.e., additional stages of measurement done on the ancilla qubits entangled with data qubits, which detect whether errors have occurred \cite{Nielsen_Chuang_2010}.

\myparagraph{Transpilation}
Now we need to adapt the circuit to the target device~\cite{maronese2021quantumcompiling}. Transpilation involves multiple stages \cite{kaya_2024, amy_gheorghiu_2020, IBM_qiskit_transpiler}: Firstly, we map the logical qubits of the circuit to physical qubits of the device on which they will be executed \cite{li2019tacklingqubitmappingproblem}. To make two-qubit gate execution possible, qubits involved in them need to be placed on physically adjacent hardware. However, as that is not always possible, we follow this stage with routing, during which we insert a gate sequence (typically an additional swapping operation) to move together the quantum states that must interact \cite{qecc_synth_2025}. To keep overhead low, we aim for routing sequences that minimize the number of movements. We then translate each gate in the circuit into an equivalent operation from the elementary set supported by the target device \cite{ge2024quantumcircuitsynthesiscompilation}, as we illustrate in Fig.~\ref{fig:qec_pipeline_example}(b). Since mapping the large connection graph, inserting SWAP operations, and decomposing unsupported gates produce substantial overhead, we interleave those mandatory stages with optimization passes, which aggressively prune the length and the gate count of the circuit~\cite{Karuppasamy_2025}, e.g., by removing mutually cancelling gates.

\myparagraph{Execution}
The promise of quantum computing \cite{shor_1995, grover1996fastquantummechanicalalgorithm} has spurred the development of diverse hardware platforms \cite{Huang2020, Henriet_2020, trapped_bruzewicz_2019}, each with distinct trade-offs in coherence times, gate speed, connectivity, and noise profiles, as summarized in Tab.~\ref{tab:technologies}. Superconducting devices, the most common platform, offer fast gates but are constrained by limited qubit lifetimes and fixed, sparse connectivity, making them particularly susceptible to idling errors, crosstalk between neighboring qubits, and more gate errors introduced by additional SWAP operations \cite{Huang2020, Willow2025}. In contrast, trapped-ion and neutral atoms devices provide longer coherence and can achieve dynamic, all-to-all connectivity via mid-circuit qubit movement, but are affected by errors arising from qubit transport and imperfect control lasers \cite{trapped_bruzewicz_2019, Wang2021, radnaev2025universalneutralatomquantumcomputer, Moses_2023, Bluvstein_2023}. Each platform's unique error characteristics  are a critical factor for the effectiveness of QEC~\cite{qpandora_chatterjee_2025}.

During circuit execution, those errors can prevent us from obtaining meaningful results. To handle them, we periodically take syndrome measurements and send the results to a classical processor, commonly an FPGA \cite{liyanage2023scalablequantumerrorcorrection} or GPU \cite{ferraz2025gpuacceleratedsyndromedecodingquantum}, where a decoder, usually tailored to the chosen QEC code, locates errors and selects the appropriate corrections, trading off precision and speed to apply them before qubits decohere~\cite{Battistel_2023}. The performance depends on the specific combination of QEC code, decoding algorithm, and hardware, especially regarding latency. This process, which we illustrate in Fig.~\ref{fig:qec_pipeline_example}(c), is essential to suppress noise and ultimately enable effective circuit execution.

\takeaway{ \label{subsec:influence_qec}
\textbf{Takeaway:}
Effective QEC performance depends on the entire quantum computing stack, as every stage from compilation to execution introduces potential overhead and errors. {\em First}, encoding and translation decisions, such as the frequency of syndrome measurements and the conversion to fault-tolerant gates, create significant qubit and gate overhead \cite{Leblond_2024, hao2025reducingtgatesunitary}. {\em Second}, the transpilation of large QEC circuits adds substantial routing overhead, with each extra gate introducing more noise that degrades the code's effectiveness \cite{peham2023depthoptimalsynthesiscliffordcircuits, Escofet_2024}. {\em Last}, the hardware and decoder are critical; a device's topology and noise profile determine a code's viability, while a slow or inaccurate decoder can introduce more errors than it corrects \cite{Benito2025comparativestudyof, bacon_surface_shor_trapped_ion_2020, surface_vs_realistic_noise_2014}.
} 

%% file: tables/technologies.tex
\begin{table}[t]
\centering
\caption{Overview of quantum technologies. {\em Quantinuum (Q.) Apollo~\cite{Quantinuum2024Roadmap} data are extrapolated from H2 device specifications \cite{QuantinuumHardwareSpecs}. For Infleqtion, reset error is included in readout error, leakage in gate error, and SPAM is approximated as measurement plus reset, as it is not directly reported~\cite{radnaev2025universalneutralatomquantumcomputer}.}}
\fontsize{8}{9}\selectfont
\begin{tabular}{|c|c|c|c|c|}
\hline
\textbf{Characteristic} & \multicolumn{2}{c|}{\textbf{Superconducting}} & \textbf{Trapped-Ion} & \textbf{Neutral Atoms} \\
\hline
Qubit Type & \multicolumn{2}{c|}{Transmon qubits \cite{Roth_2023}} & Trapped ions \cite{trapped_bruzewicz_2019} & Rydberg atoms \cite{Henriet_2020} \\
\hline
Topology & 
\multicolumn{2}{c|}{\makecell[c]{heavy-hex \cite{IBM_roadmap},\\ grid \cite{Willow2025}}} & 
\makecell[c]{race track \cite{Moses_2023},\\ grid \cite{Quantinuum2024Roadmap}} & 
\makecell[c]{2D array \cite{radnaev2025universalneutralatomquantumcomputer} \\ arbitrary 3D geometry \cite{Barredo_2018}} \\
\hline
\makecell[c]{Mid-Circuit Qubit Movement} & \multicolumn{2}{c|}{No} & Yes \cite{Moses_2023} & Yes \cite{Bluvstein_2023} \\
\hline
Coherence Time &  \multicolumn{2}{c|}{10 - 100 $\mu s$ \cite{Krantz_2019}} & $>$ 4s \cite{Wang2021, Quantinuum2024Roadmap} & 1 - 60s \cite{Wintersperger2023} \\
\hline
Operation Speed &  \multicolumn{2}{c|}{12 - 25 ns \cite{Sycamore_2019}} & 1-10 $\mu$s \cite{Schäfer2018} & 0.4 - 2 $\mu$s \cite{Wintersperger2023} \\ 
\hline
\hline
\textbf{Real Error Rates} & \textbf{Willow} \cite{Willow2025} & \textbf{Flamingo} \cite{AbuGhanem_2025} & \textbf{Q. Apollo} \cite{QuantinuumHardwareSpecs, Quantinuum2024Roadmap} & \textbf{Infleqtion} \cite{radnaev2025universalneutralatomquantumcomputer} \\
\hline
\hline
SQ & $6.2\times10^{-4}$ & $2.5\times10^{-4}$ & $8.0\times10^{-6}$ & $9.8\times10^{-4}$ \\
\hline
2Q & $2.8\times10^{-3}$ & $2.0\times10^{-3}$ & $1.4\times10^{-4}$ & $6.5\times10^{-3}$ \\
\hline
SPAM & $9.5\times10^{-3}$ & $2.5\times10^{-4}$ & $1.33\times10^{-4}$ & $4.0\times10^{-3}$ \\
\hline
Idle & $9.0\times10^{-3}$ & - & $5.3\times10^{-5}$ & - \\
\hline
Crosstalk & $5.5\times10^{-4}$ & - & $6.3\times10^{-7}$ & - \\
\hline
Leakage & $2.5\times10^{-4}$ & - & $4.3\times10^{-5}$ & - \\
\hline

\end{tabular}

\label{tab:technologies}
\end{table}

%% file: chapters/overview.tex
\input{tables/scope_coverage}
\section{Overview}
In this section, we discuss the scope of our study along with its limitations and introduce the research questions we aim to address (\S~\ref{subsec:scope}). Then, we present a systematic taxonomy of QEC codes to enable their methodical exploration (\S~\ref{sec:taxonomy}). Finally, we propose our novel benchmarking framework (\S~\ref{sec:eccentric}), which we later use for the experimental evaluation (\S~\ref{sec:hardware},  \S~\ref{sec:compilation}, \S~\ref{sec:qec_analysis}).

\subsection{Research Scope} 
\label{subsec:scope}

Each of the steps introduced in \S~\ref{sec:background} requires in-depth exploration to evaluate its impact on error correction. Taken together, these steps create a combinatorial explosion of factors, making exhaustive analysis infeasible within a single study. To manage this complexity, we slightly narrow the scope. We exclude the construction of FT logical gates, as, to the best of our knowledge, the only existing framework supporting them is limited to the surface code and still under active development \cite{tqec}, and developing a general, scalable FT gate constructor would constitute a substantial standalone contribution. FT gates would require far more resources and greatly increase circuit depth, resulting in higher logical error rates. We also omit classical decoder latency, which depends strongly on the code, decoder, and hardware platform, making any arbitrary artificial value unsuitable for general analysis. Since real-time decoding is planned to operate under 1 $\mu s$, the resulting decoherence errors in a real system would be only marginally higher than those observed here \cite{Battistel_2023}. We further discuss both limitations in \S~\ref{sec:limitations}. Additionally, because circuit optimization techniques generally improve performance by reducing gate overhead, we treat them as a secondary consideration during compilation.

Even with these restrictions, the remaining scope remains substantial. To approach it systematically, we organize our work around four key dimensions: QEC codes, quantum compilation, quantum device noise models, and quantum device topology. Based on these dimensions, we formulate nine research questions, summarized in Tab.~\ref{tab:research_questions}, which probe their impact on QEC effectiveness. We select these questions to ensure comprehensive coverage while focusing on practical aspects of compilation and execution, reflecting current development directions and targeting well-known bottlenecks (e.g., limited device size or transpilation overhead). By addressing them experimentally, we aim to generate practical insights that guide real-world QEC implementations.

\input{chapters/qecc_taxonomy}
\input{chapters/framework_design}

%% file: tables/scope_coverage.tex
\begin{table}[t]
\centering
\caption{Research questions we pursue in this work, their relations to the considered problem dimensions, and the sections where they are discussed.}
\vspace{-3pt}
\fontsize{8}{9}\selectfont
\begin{tabular}{!{\vrule width 0.4pt}c!{\vrule width 0.4pt}>{\raggedright\arraybackslash}p{6cm}!{\vrule width 0.4pt}c!{\vrule width 0.4pt}c!{\vrule width 0.4pt}c!{\vrule width 0.4pt}c!{\vrule width 0.4pt}}
\hhline{|------|}
\textbf{Sec.} & \textbf{Research Question} & \cellcolor{blue!15}\textbf{Codes} & \cellcolor{red!15}\textbf{Topologies} & \cellcolor{green!15}\textbf{Noise} & \cellcolor{yellow!15}\textbf{Compilation} \\
\hhline{|------|}
\multirow{5}{*}{\S\ref{sec:hardware}} 
& \textbf{RQ\#1}: Will the effectiveness of the QEC codes noticeably change when their distance expands? 
  & \cellcolor{blue!15}\cmark 
  & \cellcolor{red!15}\cmark 
  & \cellcolor{green!15}\xmark 
  & \cellcolor{yellow!15}\xmark \\ \hhline{|~-----|}
& \textbf{RQ\#2}: Does higher qubit connectivity always improve the effectiveness of QEC codes? 
  & \cellcolor{blue!15}\cmark 
  & \cellcolor{red!15}\cmark 
  & \cellcolor{green!15}\xmark 
  & \cellcolor{yellow!15}\cmark \\ \hhline{|~-----|}
& \textbf{RQ\#3}:
Does qubit quality variance critically impact logical error rates beyond the effect of the mean?
  & \cellcolor{blue!15}\xmark 
  & \cellcolor{red!15}\xmark 
  & \cellcolor{green!15}\cmark 
  & \cellcolor{yellow!15}\cmark \\ \hhline{|~-----|}
& \textbf{RQ\#4}:
Does a QEC code's performance scaling differ between single-QPU and distributed execution?
  & \cellcolor{blue!15}\cmark 
  & \cellcolor{red!15}\cmark 
  & \cellcolor{green!15}\xmark 
  & \cellcolor{yellow!15}\xmark \\ \hhline{|~-----|}
&\textbf{RQ\#5}:
Which quantum technology provides favorable error rates and hardware features for effective QEC?
  & \cellcolor{blue!15}\cmark 
  & \cellcolor{red!15}\cmark 
  & \cellcolor{green!15}\cmark 
  & \cellcolor{yellow!15}\xmark \\
\hline
\multirow{2}{*}{\S\ref{sec:compilation}} 
& \textbf{RQ\#6}: How do mapping and routing compilation stages influence the effectiveness of QEC codes? 
  & \cellcolor{blue!15}\cmark 
  & \cellcolor{red!15}\xmark 
  & \cellcolor{green!15}\xmark 
  & \cellcolor{yellow!15}\cmark \\ \hhline{|~-----|}
& \textbf{RQ\#7}: How does the translation stage influence the effectiveness of QEC codes? 
  & \cellcolor{blue!15}\cmark 
  & \cellcolor{red!15}\xmark 
  & \cellcolor{green!15}\xmark 
  & \cellcolor{yellow!15}\cmark \\
\hline
\multirow{2}{*}{\S\ref{sec:qec_analysis}} 
& \textbf{RQ\#8}:
Which decoders achieve the best performance across QEC code families?
  & \cellcolor{blue!15}\cmark 
  & \cellcolor{red!15}\xmark 
  & \cellcolor{green!15}\cmark 
  & \cellcolor{yellow!15}\xmark \\ \hhline{|~-----|}
& \textbf{RQ\#9}:
Is applying QEC always beneficial, or can the QEC overhead become a net source of error?
  & \cellcolor{blue!15}\cmark 
  & \cellcolor{red!15}\xmark 
  & \cellcolor{green!15}\cmark 
  & \cellcolor{yellow!15}\xmark \\
\hline
\end{tabular}
\label{tab:research_questions}
\end{table}

%% file: chapters/qecc_taxonomy.tex
\subsection{Quantum Error Correction Codes Taxonomy}
\label{sec:taxonomy}

To enable a fair and systematic evaluation across the vast landscape of QEC codes, we first introduce a taxonomy to structure our analysis. As shown in Tab.~\ref{tab:code_categories}, we organize codes into four main families based on their core principles: subsystem stabilizer, QLDPC, concatenated, and topological codes. We describe each family through its core principles and highlight important variations where relevant:

\input{tables/taxonomy_table}

\myparagraph{Stabilizer codes}
Stabilizer codes are built upon a group of mutually commuting Pauli operators \( \mathcal{S} \), where a state \( \ket{\psi} \) satisfies \( S\ket{\psi} = \ket{\psi} \) for all \( S \in \mathcal{S} \) \cite{gottesman1997stabilizercodesquantumerror}. Errors are detected through syndrome extraction: data qubits are entangled with ancillas prepared in known states, which are then measured to reveal occurring errors \cite{Roffe_2019}. Their compatibility with efficient classical simulation has made stabilizer codes the foundation of most QEC schemes.

To describe specific code instances of stabilizer codes, we use the standard $[[n,k,d]]$ notation, where $n$ is the number of physical qubits used to encode $k$ logical qubits, and $d$ is the code distance, which is the minimal number of physical qubits that must be corrupted before an error becomes undetectable \cite{Roffe_2019}. Notably, $n$ does not include ancilla qubits required for syndrome measurement.

\myparagraph{Subsystem stabilizer codes}
Subsystem codes generalize stabilizer codes by encoding logical information into a subsystem instead of a subspace, introducing additional degrees of freedom, called gauge qubits, which do not influence the logical information and can be viewed as non-essential bits \cite{aly2006subsystemcodes}. In practice, attention centers on the \textit{subsystem stabilizer} subclass, which preserves the stabilizer formalism for efficient simulation while using gauge operators to replace some stabilizers with lower-weight checks, simplifying syndrome extraction \cite{Higgott_2021}. An example of a subsystem stabilizer code is the \textit{Bacon-Shor code} \cite{Bacon_2006}. Defined on a rectangular lattice of size \(m_1 \times m_2\), it has parameters \([[m_1 m_2, 1, \min(m_1, m_2)]]\) \cite{egan2021faulttolerantoperationquantumerrorcorrection}, and is particularly well-suited for trapped-ion devices, where under realistic noise models it can outperform the state-of-the art code with larger distance \cite{ion_traps_2020}.

\myparagraph{Topological codes}  
Topological codes encode information in global features of a lattice, making them intrinsically robust against local errors \cite{Kitaev_2003, bombin2013introductiontopologicalquantumcodes}. Their local stabilizers correspond to lattice features like vertices or plaquettes, and their distance grows with the lattice size. Among topological codes, \textit{surface codes} \cite{Fowler_2012} are the most studied and considered as state-of-the-art \cite{PhysRevLett.129.030501}, typically defined on a 2D square lattice. A prominent variant, the \textit{rotated surface code}, reduces the physical qubits overhead \cite{Horsman_2012} while preserving the code’s core properties. Another key subclass is \textit{color codes} \cite{Bombin_2006}, defined on lattices with colored plaquettes so that no two adjacent plaquettes share the same color, with each color corresponding to a subset of stabilizers. A notable example is the \textit{triangular color code}, built on a 6.6.6 hexagonal lattice with three plaquette colors, which combines strong error suppression with a versatile set of low-overhead fault-tolerant gates, making it a promising candidate for scalable quantum computing \cite{bombin2013introductiontopologicalquantumcodes, lacroix2024scalinglogiccolorcode}.

\myparagraph{Quantum low-density parity-check (QLDPC) codes}  
QLDPC codes are stabilizer codes with sparse parity-check matrices, meaning that each stabilizer involves only a small subset of qubits~\cite{PRXQuantum.2.040101}. This sparsity reduces measurement complexity but at the cost of long-range connectivity. The key advantage of QLDPC codes lies in their efficient scaling, maintaining a constant logical-to-physical qubit ratio (encoding rate) as the code distance grows. A prominent subclass of QLDPC codes is \textit{bivariate bicycle (BB) codes} \cite{Bravyi2024}, which are constructed from bivariate polynomials with regular stabilizer patterns, preserving a good encoding rate while achieving even higher physical error rates they can tolerate (error thresholds). A notable example is the \textit{gross code} $[[144,12,12]]$, which encodes 12 logical qubits using 288 physical qubits, achieving performance comparable to a surface code that would require thousands of qubits \cite{Bravyi2024}.

\myparagraph{Concatenated codes}
Concatenated codes encode quantum information by nesting one code within another \cite{knill1996concatenatedquantumcodes}. Specifically, an outer code encodes the logical qubits, and each physical qubit of this code is further encoded using an inner code. This layered structure allows flexible combinations of codes to improve error correction. A notable example of a concatenated code is the \textit{concatenated Steane code} \cite{Pato_2024}, denoted as $[[7^m,1,3^m]]$, able to achieve even lower logical error rates than the triangular color code. It is constructed by recursively concatenating the standard Steane code $[[7,1,3]]$ \cite{Steane_1996} with itself $m-1$ times.  However, a common drawback of concatenated codes is that their encoding rate decreases exponentially with the concatenation level.

\input{figures_tex/codes}

\myparagraph{Architecture-specific codes}
Additionally, although we do not distinguish them as a family, some QEC codes are designed to combine different characteristics to meet the constraints of specific quantum hardware. A prominent example is the \textit{heavy-hexagon code}, introduced as a sparse alternative to the surface code \cite{Chamberland_2020} and developed for the heavy-hex topology (e.g., IBM Heron \cite{AbuGhanem_2025}). It blends topological features, using a lattice-based layout, with subsystem stabilizer properties, leveraging gauge operators to reduce qubit connectivity.

For our study, we select representative codes from each major family: the Bacon-Shor, rotated surface, triangular color, gross, and concatenated Steane codes, each chosen as a high-performing exemplar of its category. We specifically include both the state-of-the-art surface code and the color code, as the latter has promising properties for practical implementations. Additionally, we include the heavy-hex code to test how a topology-specific code performs under general conditions. The structures of these codes are illustrated in Fig.~\ref{fig:all_codes}.

%% file: tables/taxonomy_table.tex
\begin{wraptable}{r}{0.56\textwidth}
\centering
\caption{Taxonomy of main QEC code families.}
\label{tab:code_categories}
\fontsize{7}{8}\selectfont
{
\renewcommand{\arraystretch}{1.1}
\begin{tabular}{|c|c|c|c|}
    \hline
    \multicolumn{4}{|c|}{\textbf{Subsystem}} \\
    \hline
    \multicolumn{4}{|c|}{\textbf{Stabilizer}} \\
    \hline
    \textbf{Concatenated} & \textbf{QLDPC} & \textbf{Topological} & \textbf{\makecell{Subsystem\\Stabilizer}} \\
    \hline
    Concatenated Steane & BB (Gross) & Surface, Color & Bacon-Shor \\
    \phantom{Heavy-hex} & & \multicolumn{2}{c|}{Heavy-hex} \\
    \hline
\end{tabular}
}
\end{wraptable}

%% file: figures_tex/codes.tex
\begin{figure}[t]
    \centering
    \begin{subfigure}[b]{0.3\textwidth}
        \centering
        \includegraphics[scale=0.2]{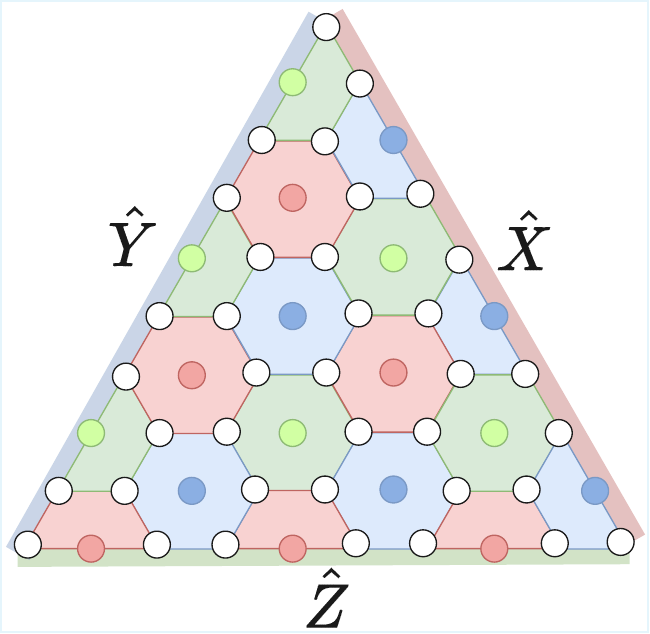}
        \caption{Triangular color code}
        \label{fig:triangular}
    \end{subfigure}
    \hfill
    \begin{subfigure}[b]{0.3\textwidth}
        \centering
        \includegraphics[scale=0.08]{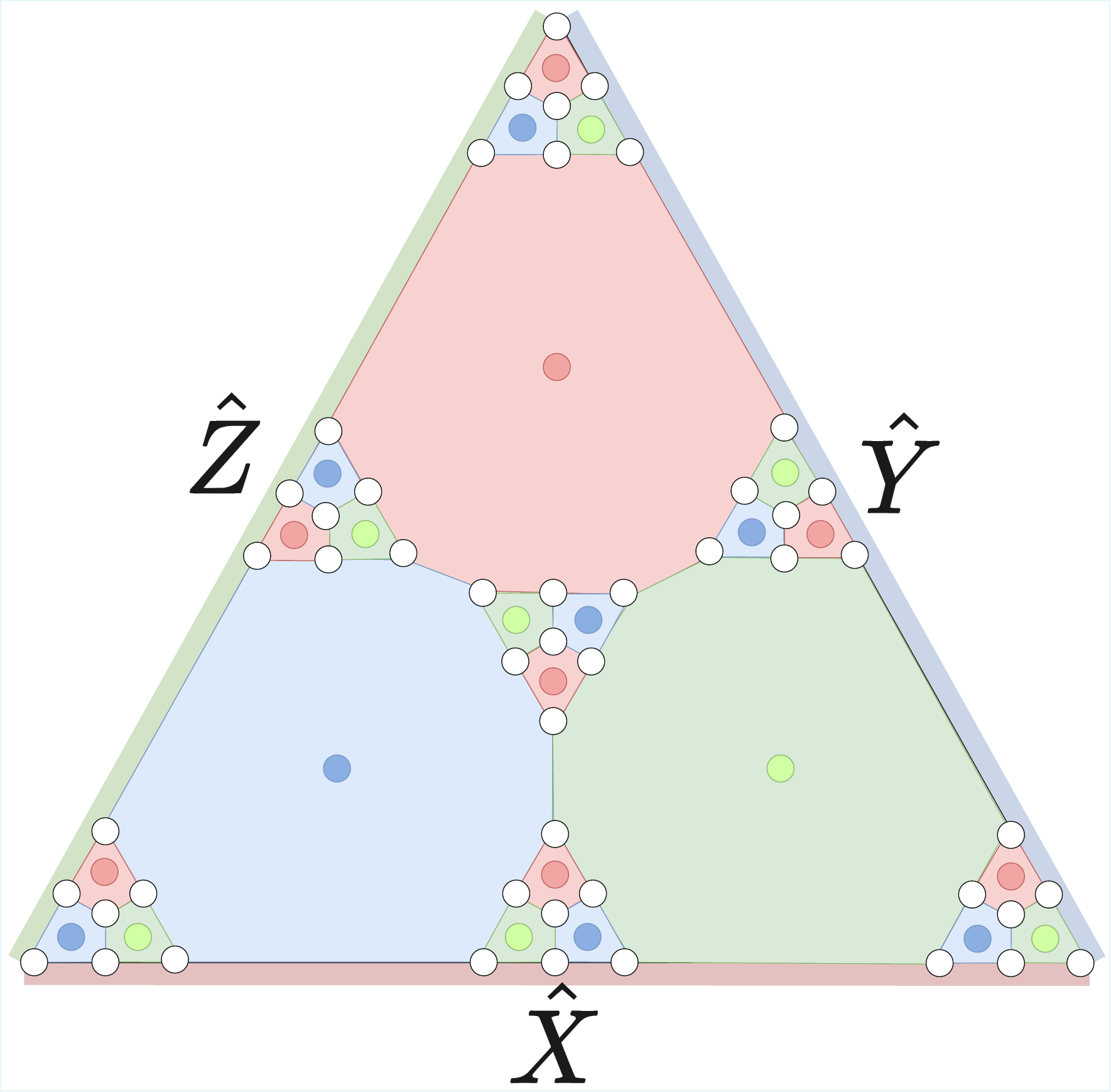}
        \caption{Concatenated Steane code}
        \label{fig:concat_steane}
    \end{subfigure}
    \hfill
    \begin{subfigure}[b]{0.3\textwidth}
        \centering
        \includegraphics[scale=0.09]{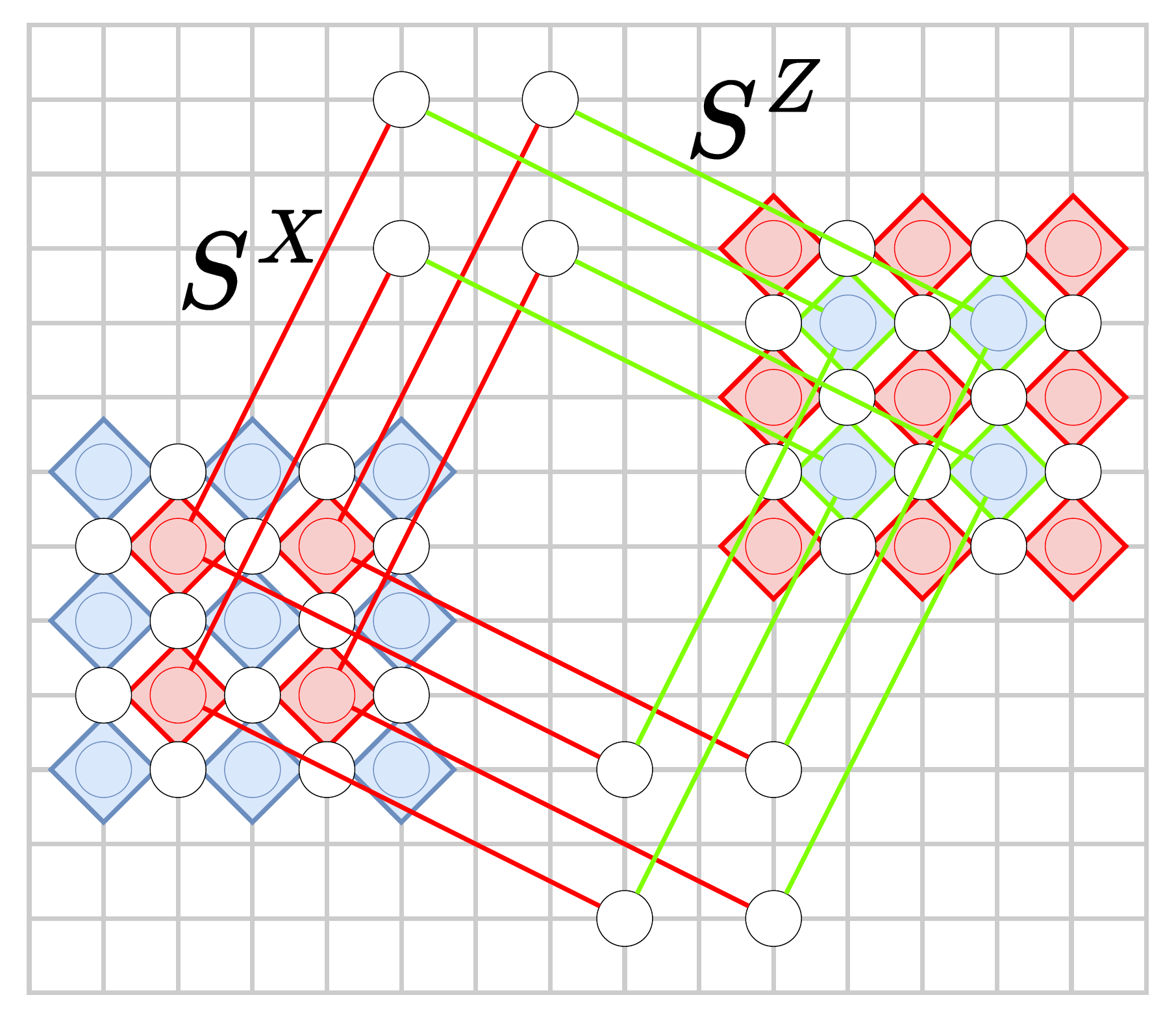}
        \caption{Gross code}
        \label{fig:gross}
    \end{subfigure}

    \begin{subfigure}[b]{0.3\textwidth}
        \centering
        \includegraphics[scale=0.045]{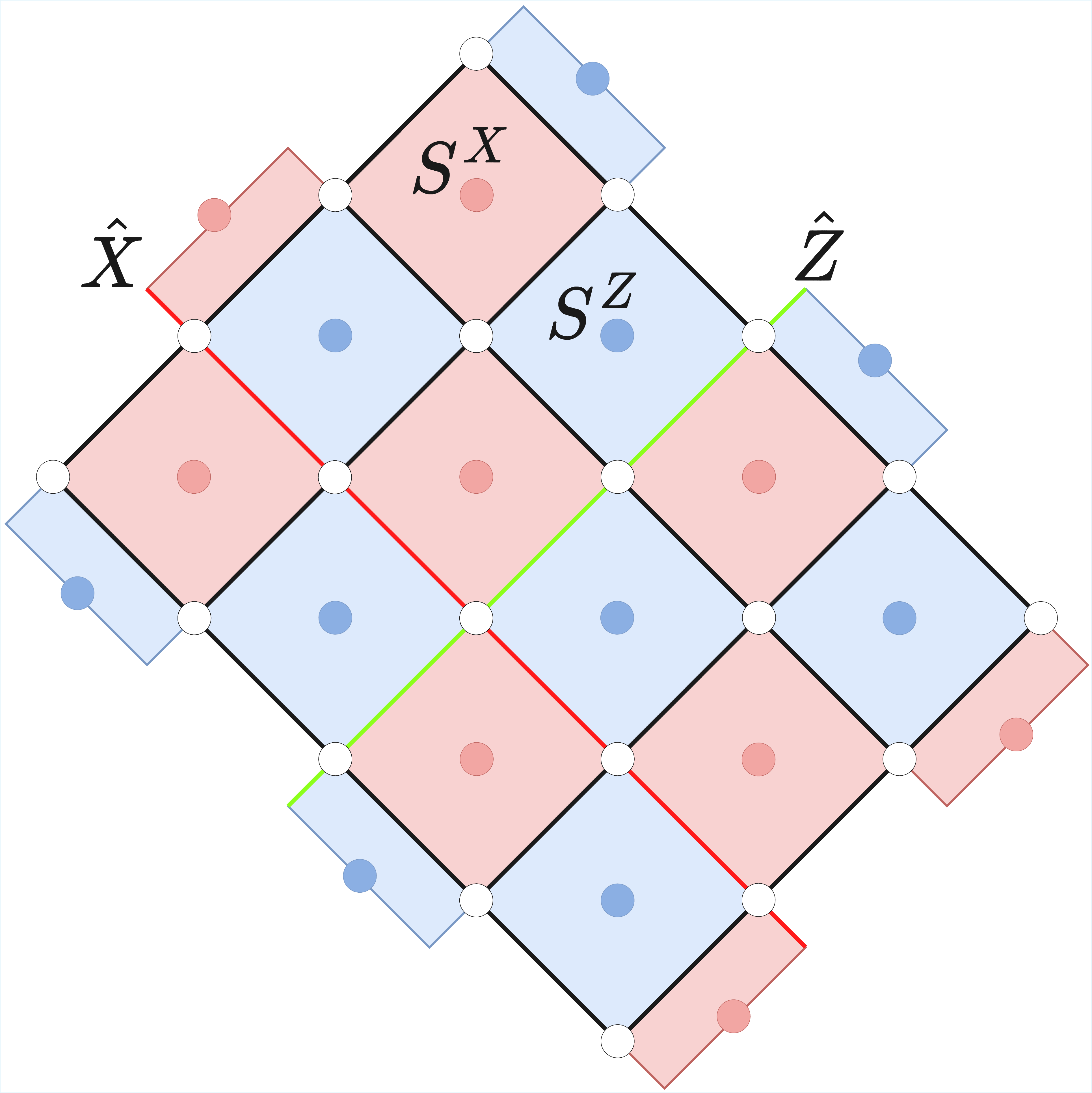}
        \caption{Surface code}
        \label{fig:surface}
    \end{subfigure}
    \hfill
    \hfill
    \begin{subfigure}[b]{0.3\textwidth}
        \centering
        \includegraphics[scale=0.12]{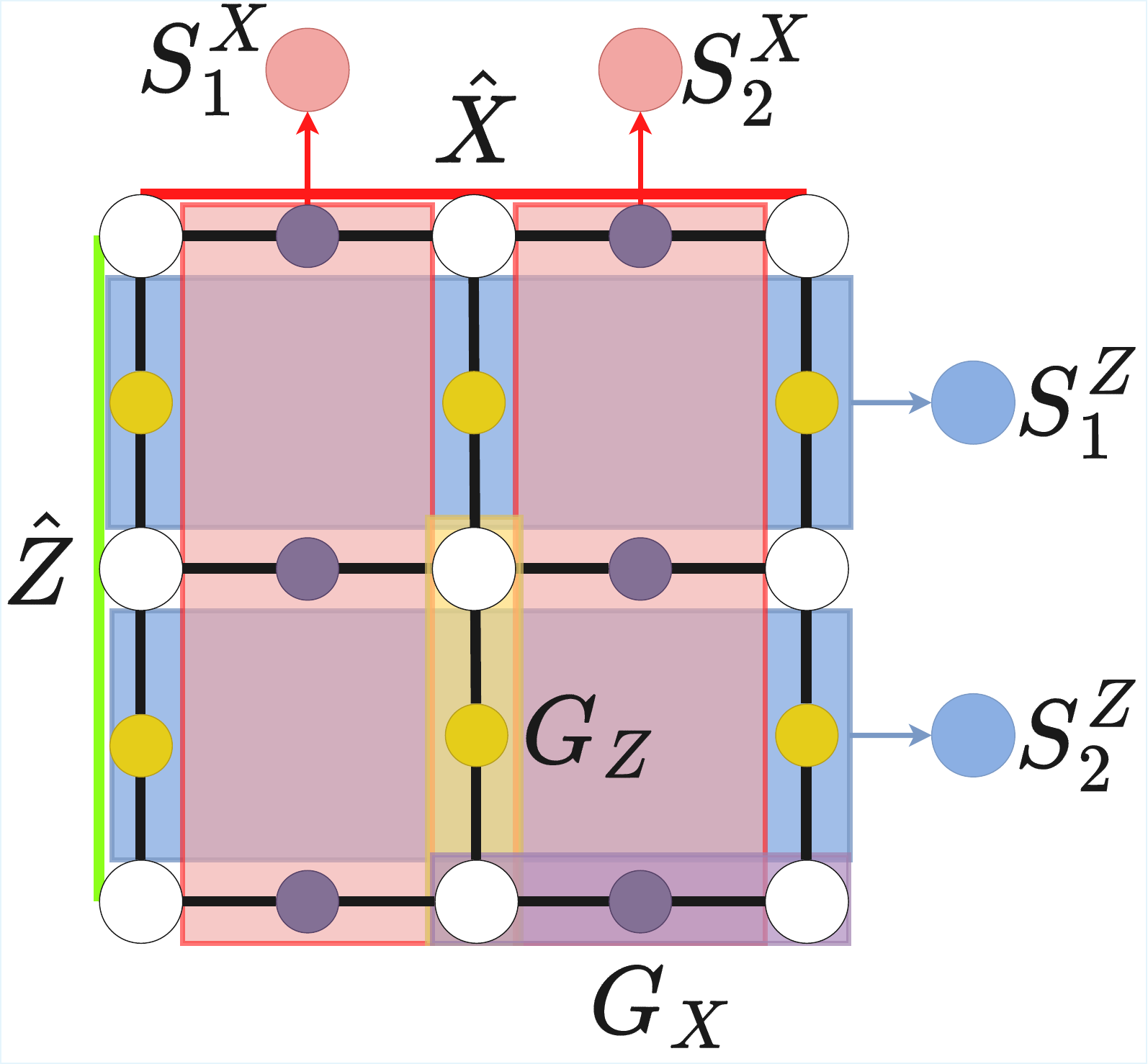}
        \caption{Bacon-Shor code}
        \label{fig:bacon_shor}
    \end{subfigure}
    \hfill
    \begin{subfigure}[b]{0.3\textwidth}
        \centering
        \includegraphics[scale=0.045]{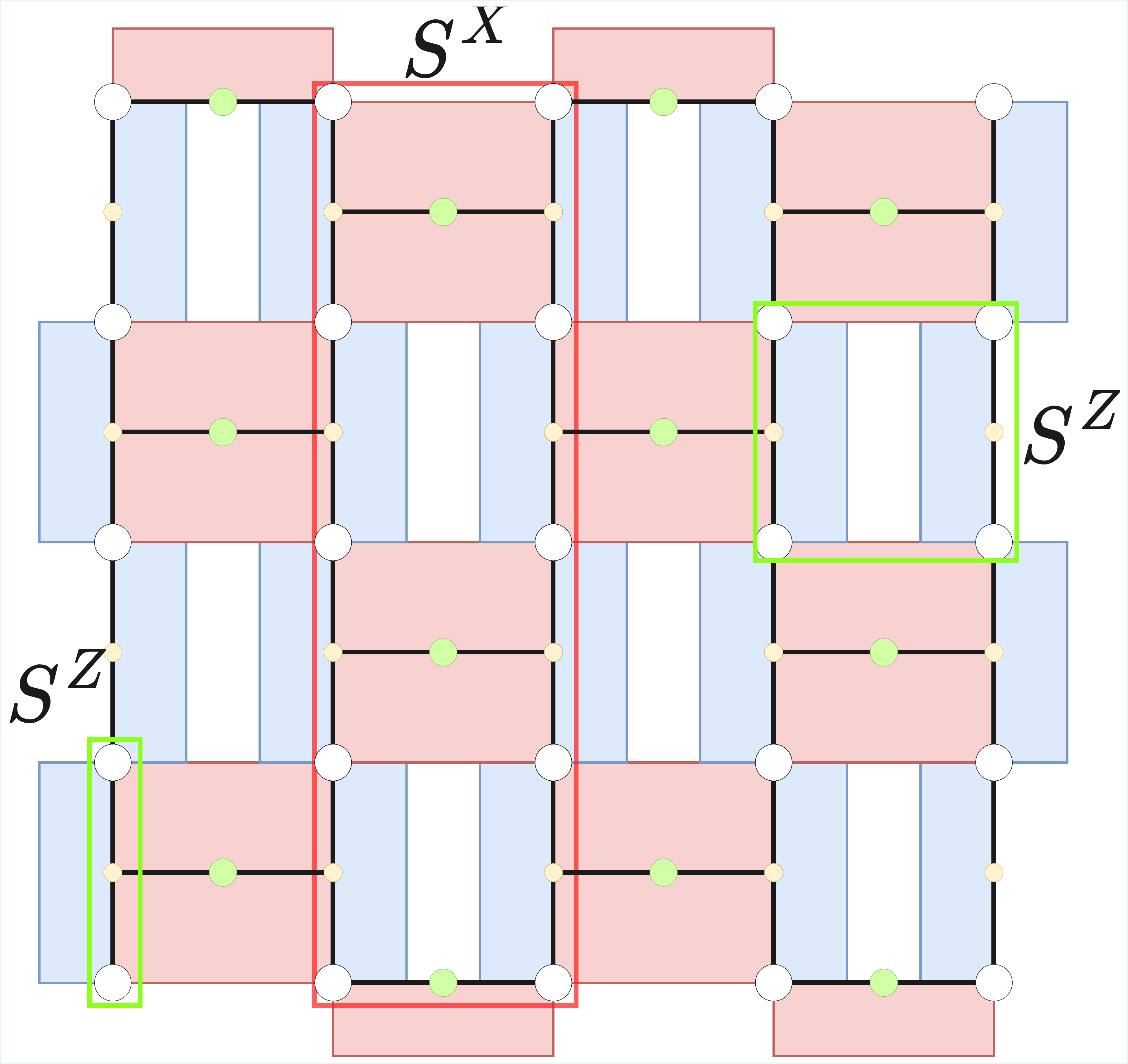}
        \caption{Heavy-Hexagon code}
        \label{fig:heavy_hex}
    \end{subfigure}

    \caption{Visualization of the six QEC codes explored in this work. {\em White dots represent data qubits, yellow and purple dots indicate gauge qubits, and remaining dots correspond to additional ancilla qubits. $\hat{X}$, $\hat{Y}$, and $\hat{Z}$ symbolize logical operators, $G_x$ and $G_z$ gauge operators. Field colors indicate stabilizers $S^x$ and $S^z$.}}

    \label{fig:all_codes}
\end{figure}

%% file: chapters/framework_design.tex
\subsection{The \projectname{} Framework}
\label{sec:eccentric}

To experimentally explore our research questions while systematically exploring the important parameters, we now require a high-performance and modular framework that incorporates different compilation stages, topology constraints, and detailed noise models. However, existing tools either support detailed compilation and execution (e.g., Qiskit \cite{javadiabhari2024quantumcomputingqiskit}) or are fast and memory-efficient enough for QEC simulation (e.g., Stim \cite{framework_stim_2021}), with no single framework fully meeting our needs. To address this gap, we develop \projectname{}, our dedicated, fully modular, extensible framework for systematic evaluation of the QEC codes, which we present in Fig.~\ref{fig:setup_general}.

\myparagraph{Encoded quantum memory generation}  
We first use pre-existing libraries \cite{framework_stim_2021, gong2024lowlatencyiterativedecodingqldpc, lee2025color, Pato_2024, gidney2023baconthreshold, heavyhexdemo} to generate quantum memory circuits, i.e., circuits representing one or more logical qubits encoded using our selected codes, not supporting FT logical gates, as discussed in \S~\ref{sec:limitations}. Generation allows arbitrary error-correction rounds and flexible code distance. For the surface, color, Bacon–Shor, and heavy-hex codes, we only generate odd-distance instances to avoid overhead or complications from majority-voting syndrome extraction \cite{Fowler_2012, Brooks_2013, Chamberland_2020_triangle}, while the gross code has a fixed distance of 12 \cite{Bravyi2024}. We further extend the Steane code with an additional concatenation layer to produce a $[[343,1,27]]$ code and provide it at distances 3, 9, and 27.
We generate all circuits in the Stim format \cite{framework_stim_2021} and subsequently translate them into Qiskit \cite{javadiabhari2024quantumcomputingqiskit} to enable execution.

\input{figures_tex/pipeline}

\input{plots/hardware_comparison_noise/hardware_noise_comparison}

\myparagraph{Translation}  
Next, circuits are transpiled to the native gate sets of target devices using high-performance translators, including Qiskit \cite{javadiabhari2024quantumcomputingqiskit}, TKET \cite{Sivarajah_2020_tket}, and BQSKit \cite{doecode_58510_bqskit} translators, selected for their performance \cite{benchpress_2025}. By default, we only use the translation functionalities of these SDKs. For full simulation, the circuits must conform to the gate set supported by Stim, so we perform an additional translation during the mapping and routing stage to guarantee that any gates not supported by Stim are correctly converted before execution.

\myparagraph{Mapping and routing}
We define all backends as custom artificial devices based on Qiskit's BackendV2 \cite{javadiabhari2024quantumcomputingqiskit}, specifying their topology and, where applicable, qubit properties, such as $T_1$ and $T_2$ times. We support both synthetic topologies and models of real devices: Google Willow \cite{Willow2025}, Quantinuum Apollo \cite{Quantinuum2024Roadmap}, Infleqtion \cite{radnaev2025universalneutralatomquantumcomputer}, IBM Flamingo \cite{IBM_roadmap}, and IBM Nighthawk \cite{IBM_roadmap_2025}. To enable execution of codes, we scale some devices up to reflect mid-term projections, such as Google Willow (3$\times$, 315 qubits) and Infleqtion (16$\times$, 384 qubits), while limiting others like Apollo to 768 qubits for simulation feasibility.

For devices supporting mid-circuit movement, we model shuttling by adding temporary, full-connectivity links with distinct durations and error rates to represent a qubit moving to its target for a remote gate and then returning. We then use the Qiskit Transpiler \cite{IBM_qiskit_transpiler} to determine the initial layout and perform routing while also exploring all available parameters: the initial layout strategies ("Trivial," "Dense," "Lookahead," and "Sabre") and routing algorithms ("Basic," "Stochastic," and "Sabre"). We also implement a qubit tracker to maintain the mapping between physical and logical qubits throughout the circuits, to ensure accurate noise modeling throughout this process.

\myparagraph{Representation change}  
To enable efficient stabilizer simulation, we translate the circuits into the Stim format. Since Qiskit circuits do not include Stim-specific gates (e.g., MRX), we represent these operations as sequences of standard gates in the resulting Stim circuits. We verify the correctness of this translation both manually and automatically across a range of code samples.

\myparagraph{Noise addition}  
After translation, we inject noise gate-by-gate to ensure compatibility with our stabilizer simulation. Our framework supports \circled{1} single- and two-qubit gate errors, \circled{2} idle errors (modeled either with a constant probability or derived from qubit-specific $T_1$/$T_2$ times and gate durations), \circled{3} leakage with propagation, \circled{4} crosstalk, \circled{5} reset errors, \circled{6} readout errors, \circled{7} qubit shuttling errors (modeled as decoherence proportional to movement time \cite{Bluvstein_2022}), and \circled{8} inter-chip gate errors, extending the approach of \cite{comparison_honeycomb_vs_surface_2021}. We provide both standard models, such as a modified SI1000 \cite{comparison_honeycomb_vs_surface_2021}, and realistic, device-specific models for platforms including Infleqtion, Google Willow, IBM Flamingo, and Quantinuum Apollo (Tab.~\ref{tab:technologies}). Fig.~\ref{fig:hardware_noise} illustrates how closely our framework can approximate an actual device.

\input{plots/num_shots/num_shots}

\myparagraph{Stabilizer simulation}
Finally, we run a stabilizer simulation using Stim \cite{framework_stim_2021}, which tracks only the stabilizers rather than full quantum states, allowing efficient, exact simulation without approximations, to sample how physical errors propagate and which syndromes they trigger. We compile these samples into detector error models \cite{derks2024designingfaulttolerantcircuitsusing} and feed them to a decoder, which infers the most likely error pattern and applies corrections. We provide two general-purpose decoders: minimum-weight perfect matching (MWPM) ~\cite{pymatching_higgott_2025} and belief propagation with order
statistic decoding (BP-OSD) ~\cite{bposd_roffe_2020}, selected for their effectiveness across all codes we study and their well-established open-source implementations. We describe the decoders and justify our choices in more detail in Appendix~\ref{sec:decoders}. The logical error rate is then estimated as the fraction of runs where errors remain uncorrected.

\myparagraph{Experimental setup}
Unless stated otherwise, we use circuits at the maximal code distance supported by the topology, with a number of error-correction rounds equal to the code distance. We run each experiment for 1000 shots, giving confidence intervals of $\approx$0.004 for logical error rates expected at error probabilities around 0.004, which are at most this value in low-error regimes, while higher rates indicate code failure. This shot count is sufficient to distinguish these regimes and assess the suitability of each code without incurring high computational cost. In Fig.~\ref{fig:shots}, we show that for the gross code, which has the most gates and therefore experiences more opportunities for errors, the logical error rates for 1000 and 10000 shots overlap, indicating that 1000 shots already provide sufficient statistical resolution across this error range.

%% file: figures_tex/pipeline.tex
\begin{figure}[t]
    \centering
    \includegraphics[scale=0.28]{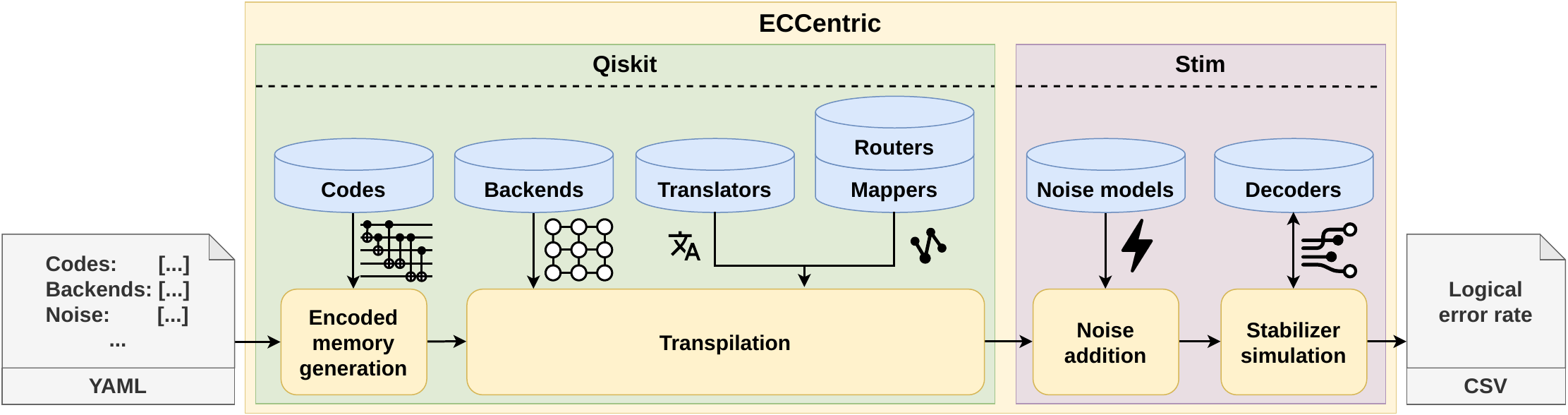}
    \caption{Architecture of the \projectname{} framework. {\em Our design of a fully modular and extensible setup, enabling systematic study of the different aspects of QEC realization.}}
    \label{fig:setup_general}
\end{figure}

%% file: plots/hardware_comparison_noise/hardware_noise_comparison.tex
\begin{wrapfigure}{r}{0.4\textwidth}
    \centering
    \includegraphics[scale=0.45]{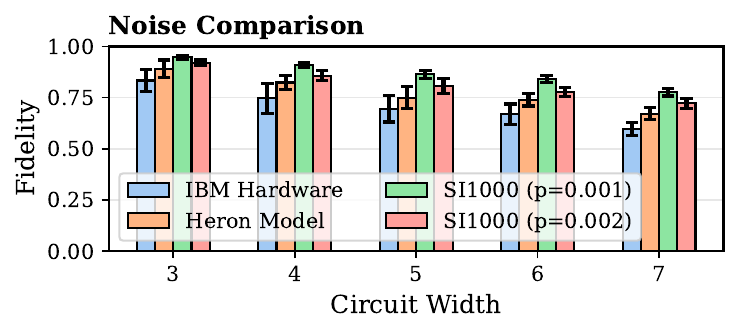}
    \caption{Comparison of real hardware, the Heron noise model, and standard artificial noise models. {\em Greater similarity to hardware indicates a more realistic model. We run 10 circuits per width with 1000 shots and report mean fidelity relative to a noiseless simulation.}}

    \label{fig:hardware_noise}
\end{wrapfigure}

%% file: plots/num_shots/num_shots.tex
\begin{wrapfigure}{r}{0.35\textwidth}
    \centering
    \includegraphics[scale=0.45]{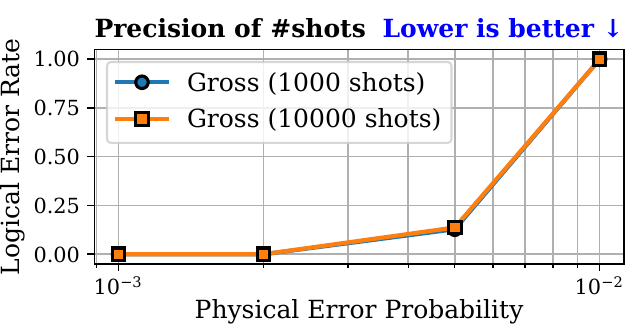}
    \caption{Effect of the number of shots on logical error rate precision.
    {\em Logical error rate of the gross code under the SI1000 noise model for 1000 and 10000 shots.}}
    \label{fig:shots}
\end{wrapfigure}

%% file: chapters/hardware_restriction_analysis.tex
\section{Hardware Restriction Analysis}
\label{sec:hardware}
\glsresetall
In this section, we present a detailed effectiveness analysis of the chosen QEC codes across the hardware-related dimensions: topology size (\S~\ref{subsec:size}), topology connectivity
(\S~\ref{subsec:connectivity}), variance of qubit quality (\S~\ref{subsec:variance}), distributed device (\S~\ref{subsec:dqc}), and different realistic quantum technologies (\S~\ref{subsec:technology}).

\subsection{Topology Size}
\label{subsec:size}

\input{tables/quantum_error_codes_comparison}

\begin{wrapfigure}{r}{0.43\textwidth}
    \centering
    \includegraphics[scale=0.35]{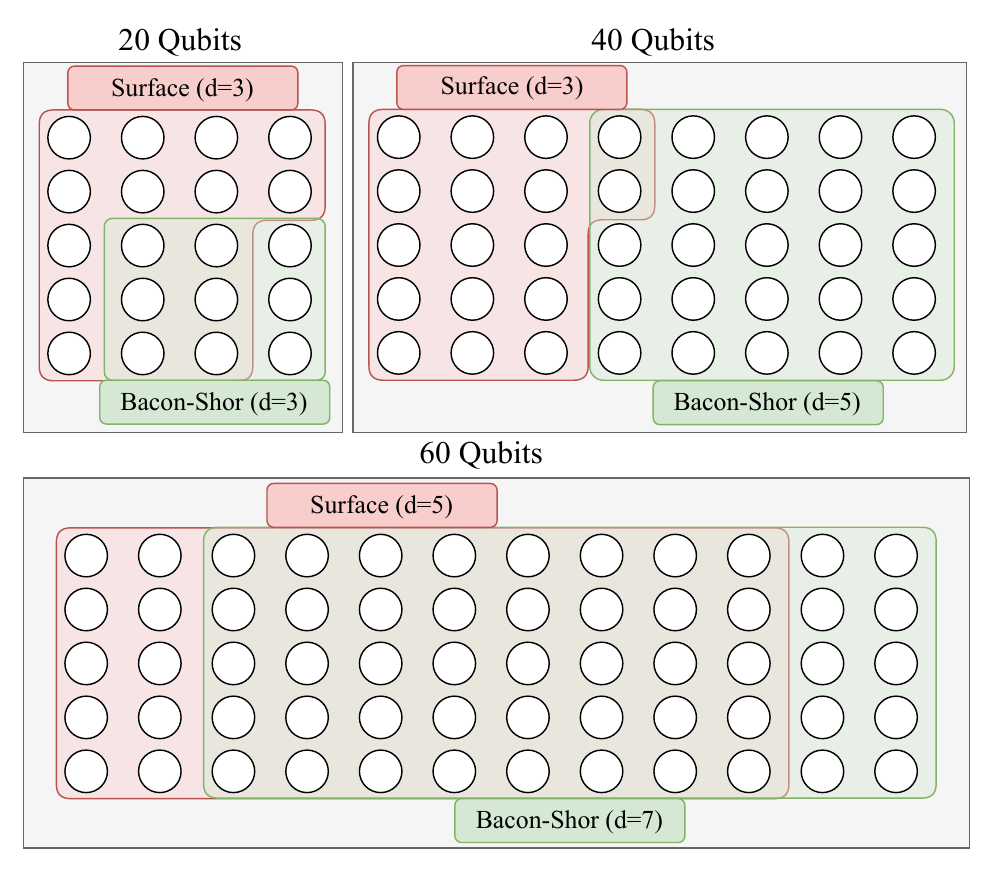}
    \caption{QEC codes possible on different device sizes. {\em An overview of the attainable distances for the surface code (in red) and the Bacon–Shor code (in green) on systems with 20, 40, and 60 qubits.}}
    \label{fig:size_example}
\end{wrapfigure}

We examine how the performance of QEC codes scales with the number of physical qubits. The scale of quantum hardware is rapidly increasing, with roadmaps projecting growth from hundreds of qubits on current devices, such as Google's Willow \cite{Willow2025} and IBM's Flamingo \cite{IBM_roadmap}, to thousands on future Quantinuum systems \cite{Quantinuum2024Roadmap}. This expansion in qubit count enables the usage of larger codes. Conventionally, a larger code distance is expected to yield superior error suppression, as it directly determines the maximum number of correctable errors \cite{Roffe_2019}. Tab.~\ref{tab:code_stats} lists the encoding rate of each code, and Fig.~\ref{fig:size_example} shows the maximum possible distance for the Surface and Bacon-Shor codes on devices with 20, 40, and 60 qubits.

\myparagraph{Research question and hypothesis}  This leads to the question: \textbf{RQ\# 1}: {\em Does a QEC code's effectiveness change as its distance increases?} Our hypothesis is that a higher distance is always more beneficial, and thus, the decision on how many qubits to use will lead to a trade-off between using the entire QPU for the best protection of one qubit or having more qubits under less severe protection. Fig.~\ref{fig:size_example} illustrates a key resource trade-off using a 60-qubit device. With the surface code, one could implement a single distance-5 patch or three distance-3 patches. With the Bacon-Shor code, the options include one distance-7 patch, two of distance-5, or six of distance-3. 

\myparagraph{Methodology} To investigate this trade-off, we simulate our selected QEC codes with 1000 shots at their maximum possible distance, corresponding to the distance number of rounds on an idealized, all-to-all connected 300-500 qubit backend. For comparison, we include the fixed-distance gross and concatenated Steane ([[49,1,9]]) codes as baselines. The simulations employ two error models: the SI1000 model and a more severe uniform error model with an error probability of 0.004.

\begin{figure}[t]
    \centering
    \includegraphics[scale=0.45]{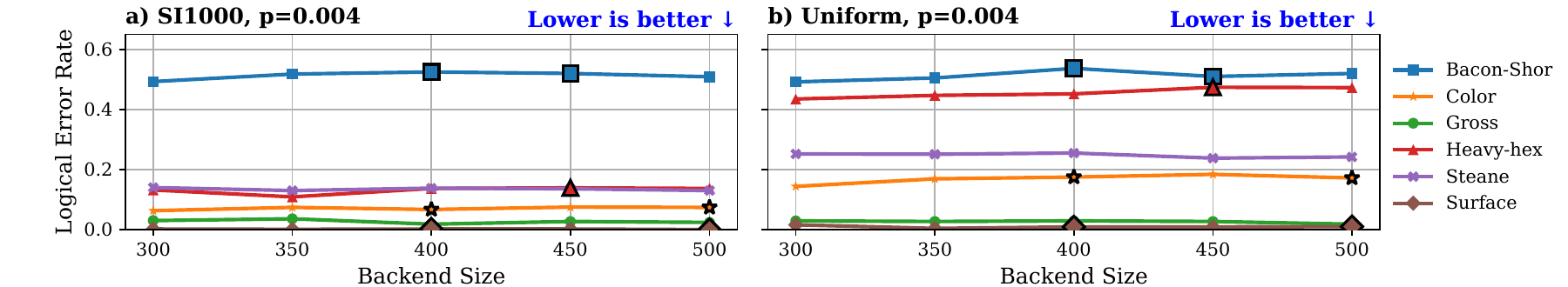}
    \caption{Logical error rates of different QEC codes at the maximal available code distance across backend sizes under two noise models of error probability 0.004. {\em Highlighted points mark an increase in the maximal achievable code distance.}}
    \label{fig:size_4}
\end{figure}

\myparagraph{Results} Our results in Fig.~\ref{fig:size_4} show that a code's distance provides surprisingly little benefit. We observe no consistent improvement with distance; a larger distance on average corresponds to a slight increase in the logical error rate. Notably, at this error probability, only surface and gross code remain effective. 

\myparagraph{Analysis}
We repeat the experiment at two additional error probabilities, 0.002 and 0.008, to examine whether the effect persists closer to and farther from the codes' thresholds. Across all four codes and both probabilities, distance variations rarely coincide with the largest deviations from the mean, confirming no consistent trend. At 0.002, all codes except Bacon-Shor perform better, achieving logical error rates of at most 0.156. Color code and gross code produce almost error rate near zero, for surface code we observe no errors. At 0.008, performance degrades across all codes; surface code holds up best under both models, though its logical error rate still exceeds 10\%.

\begin{figure}[t]
    \centering
    \includegraphics[scale=0.45]{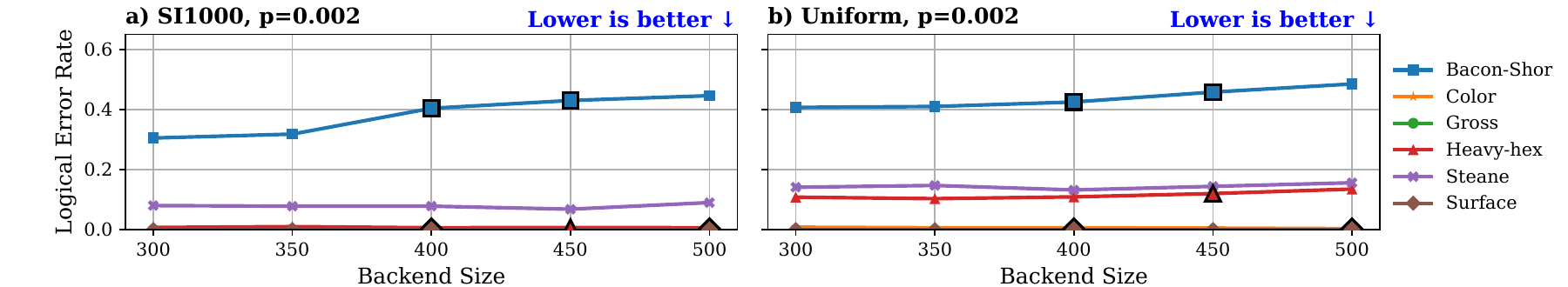}
    \caption{Logical error rates of different QEC codes at the maximal available code distance across backend sizes under two noise models of error probability 0.002. {\em Highlighted points mark an increase in the maximal achievable code distance.}}
    \label{fig:size_2}
\end{figure}
\begin{figure}[t]
    \centering
    \includegraphics[scale=0.45]{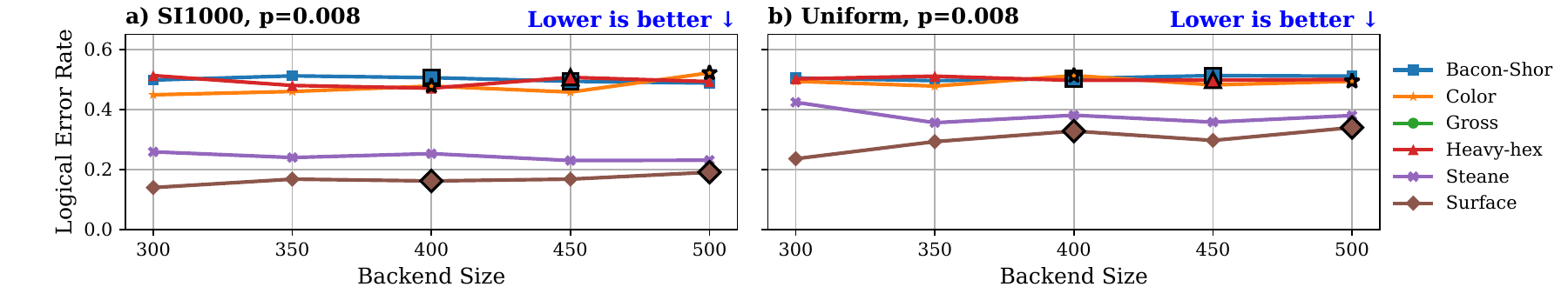}
    \caption{Logical error rates of different QEC codes at the maximal available code distance across backend sizes under two noise models of error probability 0.008. {\em Highlighted points mark an increase in the maximal achievable code distance.}}
    \label{fig:size_8}
\end{figure}

\takeaway{\textbf{Takeaway \#1:} Our experiments show that increasing code distance does not consistently improve logical performance. The change, on average, raises logical error rates by around 0.012 due to the overhead of additional qubits and gates. This suggests that once a code shows effectiveness, we can use the available device scale to run larger protected circuits instead of raising the distance, enabling more complex algorithms on mid-term hardware.}

\subsection{Topology Connectivity}
\label{subsec:connectivity}

\begin{wrapfigure}{r}{0.28\textwidth}
    \centering
    \includegraphics[scale=0.25]{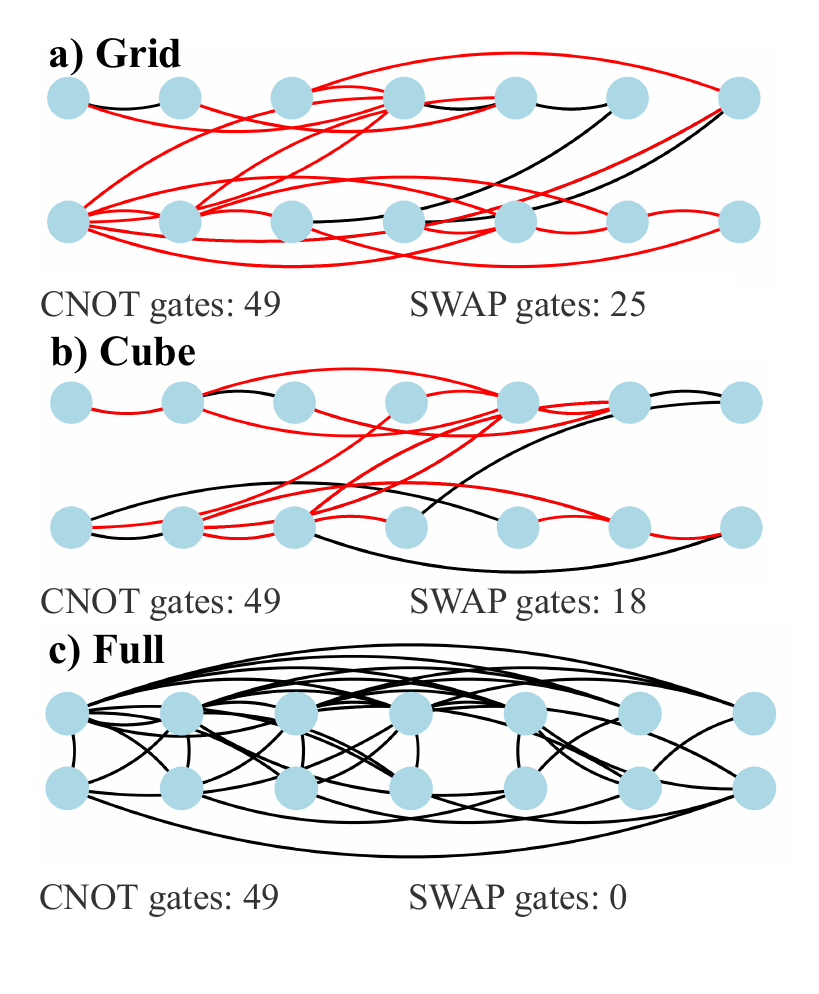}
    \caption{Connectivity graphs of Steane code $[[7,1,3]]$ circuit.
    {\em Circuit mapped and routed to various topologies. Black links indicate CNOT connections; red links indicate SWAP connections. Overlapping connections prioritize the noisier SWAP.}}
    \label{fig:connectivity_example}
\end{wrapfigure}

Qubit connectivity is a critical factor for QEC because certain hardware topologies (e.g., superconducting) require noisy SWAP gates for routing, which significantly increases error rates \cite{maronese2021quantumcompiling}. For instance, Figure~\ref{fig:connectivity_example} shows that mapping the Steane code to sparse topologies such as grids or cubes requires additional, error-inducing SWAP gates, an overhead that a fully connected architecture entirely avoids.

\myparagraph{Research question and hypothesis}  These observations raise the following  question: \textbf{RQ\# 2}: {\em Does higher qubit connectivity always improve the effectiveness of QEC codes?} We hypothesize that higher qubit connectivity reduces SWAP overheads: while a Steane code circuit requires only 49 two-qubit gates on a fully connected device, mapping it to sparser topologies with the Qiskit transpiler \cite{IBM_qiskit_transpiler} adds 54 extra CNOTs on a cube layout and 75 on a grid layout. Since two-qubit gates are a dominant error source \cite{AbuGhanem_2025, Willow2025, quantinuum_h2_datasheet_2024}, this routing overhead will degrade code performance.

\myparagraph{Methodology} To isolate and quantify the impact of connectivity, we simulate each QEC code at its maximum feasible distance on three artificial 300-qubit topologies: a 15$\times$20 grid, a 5$\times$6$\times$10 cuboid, and a fully connected graph. To minimize confounding effects from noise, we use an SI1000 model with a low error probability of 0.002.

\input{plots/eval_connectivity/connectivity}

\myparagraph{Results} The logical error rates, summarized in Fig.~\ref{fig:connectivity}(a), reveal two distinct trends. First, transitioning from a 2D grid to a 3D cuboid topology yields inconsistent benefits; while the surface code improves by 24.56\%, the concatenated Steane performs 81.48\% worse. In contrast, adopting a fully connected topology yields substantial and universal improvements, reducing logical errors by an average of 81.92\% from a grid. Notably, under full connectivity, we observe no errors for the surface code and the gross code.

\myparagraph{Analysis} To validate our findings on realistic hardware, we simulate performance on Google Willow, Quantinuum Apollo, and Infleqtion topologies. Using an SI1000 noise model with a two-qubit error probability of 0.004, the results in Fig.~\ref{fig:connectivity}(b) confirm that connectivity remains the dominant factor. Apollo, which supports shuttling, achieves an average logical error rate of 0.22, compared to 0.40 on more constrained layouts. The concatenated Steane code delivers the best overall performance with an average error of 0.13; however, we observe only a single error on the surface code on Apollo. Topology dependence is most evident in the gross code: it performs well on Apollo (0.021) but fails on the other platforms.

\takeaway{\textbf{Takeaway \#2:} 
We confirm that connectivity is a crucial factor in the effectiveness of QEC. While transitioning from a 2D grid to a 3D cuboid topology offers inconsistent benefits, moving to an idealized fully connected layout is highly effective, reducing the logical error rate by {81.92\%} on average. However, such fully connected topologies are physically unrealistic for superconducting qubits due to frequency collisions and crosstalk \cite{Zhang_2025}. Hence, realistic architectures rely on partial-connectivity mechanisms, such as qubit shuttling, which approximate full connectivity over time at the cost of additional latency and noise from qubit movement. With realistic device models, we observe that shuttle capabilities improve performance by an average of {45\%.} We find that codes exhibit varying sensitivity to these constraints and confirm that high connectivity is essential to unlock the full potential of QEC.}

\subsection{Variance of Qubit Error Probabilities}
\label{subsec:variance}

We next investigate the impact of non-uniform physical qubit quality on QEC performance. While qubits in trapped-ion and neutral atoms systems are largely uniform \cite{Saffman_2016, trapped_bruzewicz_2019}, superconducting qubits exhibit significant heterogeneity in their decoherence ($T_1$) and dephasing ($T_2$) times \cite{tannu_2019} as well as their readout errors rates \cite{tannu2019mitigating}. Consequently, relying on device-wide average error rates can be misleading, as a code's performance may be dictated by the specific, lower-quality qubits on which it is executed.

\begin{wrapfigure}{r}{0.4\textwidth}
    \centering
    \includegraphics[scale=0.4]{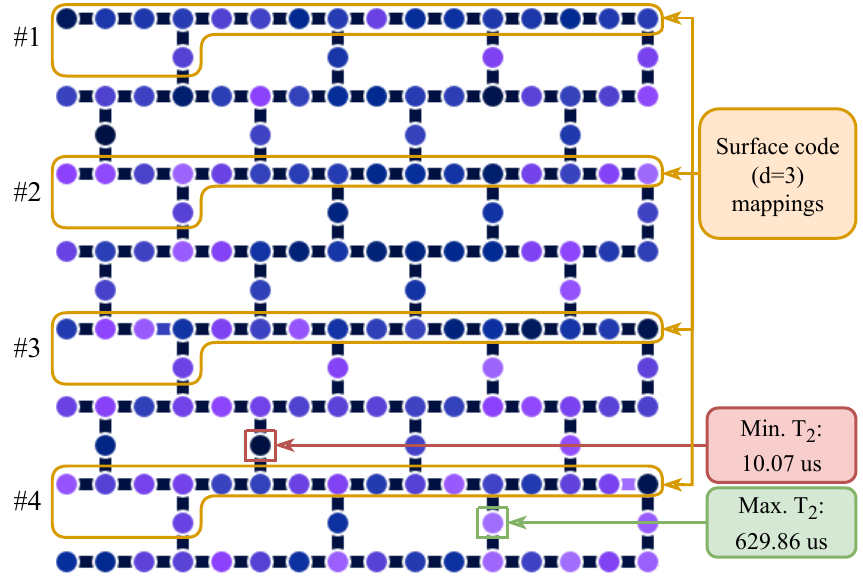}
    \caption{Qubit topology of the IBM Heron r3 \texttt{ibm\_pittsburgh} QPU.
    {\em Topology of the \texttt{ibm\_pittsburgh} device with qubit $T_2$ dephasing times ($\mu s$) shown by color intensity, based on the calibration of 2025-08-29, 07:56:08~\cite{ibm_quantum}. Example mappings of a distance-3 surface code are highlighted with orange frames.}}
    \label{fig:ibm_pittsburgh}
\end{wrapfigure}

Fig.~\ref{fig:ibm_pittsburgh} illustrates $T_2$ and gate lengths on IBM Heron’s \texttt{ibm\_pittsburgh} \cite{ibm_quantum}. We consider $T_2$ because it is shorter than $T_1$. The mean $T_2$ is 330.14 $\mu s$ and the median 336.16 $\mu s$. However, with the standard deviation of 142.84 $\mu s$, there are also qubits of significantly lower lifetimes.

\myparagraph{Research question and hypothesis}  We raise a question \textbf{RQ\# 3}: {\em Is mean qubit quality enough to
characterize expected error rates in heterogeneous
qubit devices, or does variance also play a critical role?} We hypothesize that reducing the qubit quality variance improves QEC performance by making error patterns more predictable. We show four possible mappings of a distance-3 surface code onto the \texttt{ibm\_pittsburgh} device in Fig.~\ref{fig:ibm_pittsburgh}. The choice of physical qubits significantly impacts the average lifetime of the logical qubit; the mean $T_2$ times for the mappings are 253.00 $\mu s$, 355.88 $\mu s$, 292.45 $\mu s$, and 373.63 $\mu s$.

\myparagraph{Methodology}
We simulate the QEC codes on a 399-qubit all-to-all topology using the IBM Heron noise model to isolate the effects of qubit variability from device size and connectivity constraints. Qubit quality is modeled by sampling $T_1$ and $T_2$ values from normal distributions calibrated to IBM Heron data. We explore three variance regimes with standard deviations of 0 $\mu s$ (none), 60 $\mu s$ (low), 120 $\mu s$ (medium), and 180 $\mu s$ (high), and additionally consider a tenfold prolongation of $T_1$ and $T_2$ to study the codes under reduced-noise conditions.

\myparagraph{Results}
Fig.~\ref{fig:variance}(a) shows that increasing qubit lifetime variance does not systematically affect logical error rates, with most codes varying by less than 9\%. To verify that this lack of correlation is not due to codes being already ineffective under baseline noise, we repeat the simulations with $T_1$ and $T_2$ values scaled tenfold to reduce noise, as shown in Fig.~\ref{fig:variance}(b). Even in this high-qubit-lifetime regime, no consistent trend emerges: logical error rates fluctuate non-monotonically, with pronounced oscillations, e.g., the color code's error decreases by 69.64\% from mid to high variance.

\myparagraph{Analysis} 
However, these results may stem from the relatively small overall influence of decoherence errors. To further explore this effect, we investigate the impact of varying readout errors, another commonly fluctuating source of noise. We model the readout errors using a logarithmic distribution with standard deviations of 0.0, 0.4, 0.7, and 1.0, chosen to best represent the spread of errors observed on the IBM Torino device. As shown in Fig.~\ref{fig:variance}(c), we again observe no clear correlation, with only the Steane code displaying a steady increase in logical error rates across the variation levels (from 0.30 to 0.319).

\input{plots/eval_variance/variance}

\takeaway{\textbf{Takeaway \#3:} We show that variations in qubit lifetime and readout error do not systematically affect the performance of QEC codes: no consistent correlation is observed, and the average magnitude of differences is $0.02$. This is consistent with an important feature of many QEC codes, particularly topological codes as discussed in §3.2, i.e., an immunity to local errors. As long as individual qubits remain below the code's error threshold, local variations in qubit quality have a negligible impact. Our study suggests that incorporating qubit-variance-aware or readout-variance-aware code selection and mapping is unnecessary, which can simplify compiler design.}

\subsection{Distributed Topology}
\label{subsec:dqc}

\begin{wrapfigure}{r}{0.29\textwidth}
    \centering
        \includegraphics[scale=0.44]{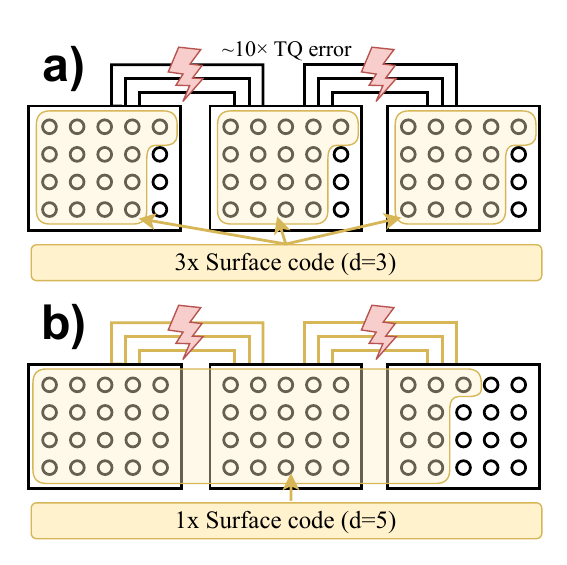}
        \caption{QEC code placement on a distributed device. 
        {\em (a) Surface code ($d{=}3$) circuits on individual QPUs. (b) Surface code ($d{=}5$) circuit spanning the full device}.}
        \label{fig:dqc_example}
\end{wrapfigure}

Next, we explore the effectiveness of QEC codes on distributed devices. Due to scalability limitations of monolithic QPUs, several companies are developing architectures in which multiple QPUs are connected via noisier and slower long-range couplers~\cite{IBM_roadmap, shapourian2025quantumdatacenterinfrastructures}. For instance, IBM Flamingo consists of three connected IBM Heron QPUs, as we show in Fig.~\ref{fig:dqc_example}. The two-qubit gate error rate for inter-QPU connections is $\sim10\times$ higher than an intra-QPU connection.

Since the code distance determines the number of corrections~\cite{Roffe_2019}, we might expect that increasing the distance across distributed QPUs should enhance fault tolerance. However, our study in \S~\ref{subsec:size}, combined with the noisy nature of inter-QPU connections, challenges this intuition.

\myparagraph{Research question and hypothesis} 
Our next question is
\textbf{RQ\#4:} {\em Does the effectiveness of a QEC code scale across the QPUs of a distributed architecture similar to when the code runs on a single QPU?} Our results in \S~\ref{subsec:variance} hint that distributing a QEC code across QPUs will not degrade its performance.
We motivate this in Fig.~\ref{fig:dqc_example}, where we scale a surface code from distance $d=3$ on a single 20-qubit QPU to $d=5$ across QPUs, which requires using a few, noisier inter-QPU links. We treat them as high-error outliers, analogous to the low-quality qubits in our previous study, and thus predict that their impact on the logical error rate will be limited.

\input{plots/eval_dqc/dqc}

\myparagraph{Methodology}
We test whether this intuition holds on the real-life IBM Flamingo topology. To do so, we compare logical error rates achieved by the QEC codes with their maximal possible distance and with distance limited by the size of a single QPU.

\myparagraph{Results} In Fig.~\ref{fig:dqc}(a), we compare logical error rates for codes executed on a single QPU versus across the full distributed topology (the gross code cannot be run on a single QPU due to its size, so we exclude it). Single-QPU execution performs substantially better, with an average improvement of 13.27\%. The surface code benefits the most (41.88\%), while the concatenated Steane code is the only code that performs better on the distributed topology, since its effective distance remains unchanged. Because high error rates render most codes ineffective, Fig.~\ref{fig:dqc}(b) shows the same comparison under the IBM Flamingo noise model with  error rates reduced by a factor of 10. Here, the single-QPU advantage grows to 55.15\%, with the surface code again gaining the most (88.63\%) and the concatenated Steane code the least (23.81\%). These results suggest that limited inter-QPU connectivity, rather than noise levels, primarily drives the performance gap.

\myparagraph{Analysis}
To test the role of connectivity, we explore the performance on IBM Nighthawk \cite{IBM_roadmap_2025}, which has the same error rates as Flamingo but higher connectivity (six neighbors per qubit) and the same quality of inter- and intra-QPU couplers. In Fig.~\ref{fig:dqc}(c), only the Steane code performs better on a single QPU under normal noise, and under scaled noise (Fig.~\ref{fig:dqc}(d)), no code does. All other codes benefit slightly from the larger distributed system, with average gains of just 0.076 (normal) and 0.072 (scaled).

\takeaway{\textbf{Takeaway \#4:} We find that QEC codes on a single QPU perform 34.21\% better on average than on distributed QPUs with limited inter-QPU connectivity. Our results show that both stronger and more numerous inter-QPU couplers are critical for achieving fault tolerance in distributed architectures.}

\subsection{Quantum Technologies}
\label{subsec:technology}

\begin{wrapfigure}{r}{0.55\textwidth}
    \centering
    \includegraphics[scale=0.25]{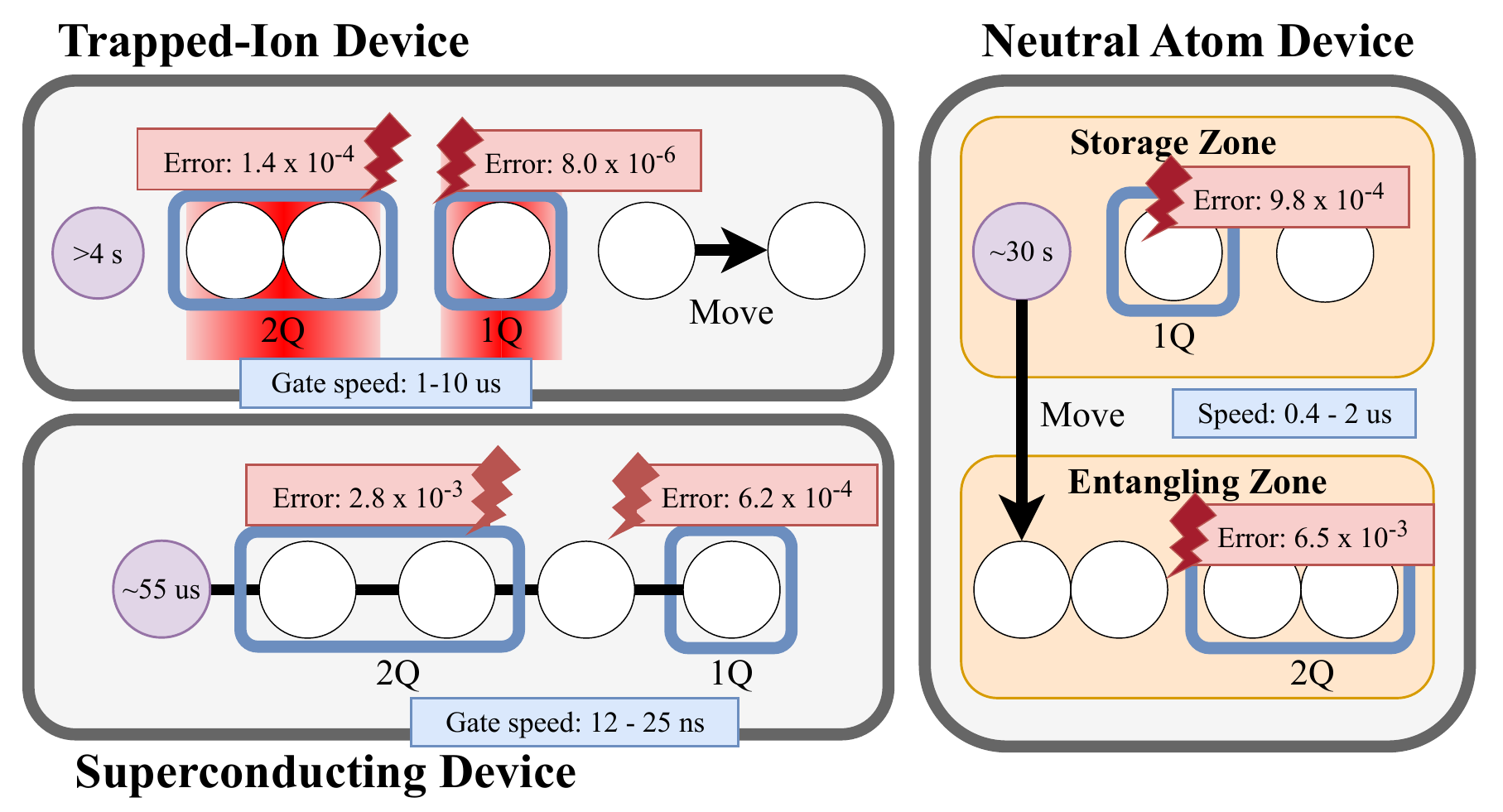}
    \caption{Overview of quantum technologies. 
    {\em Gate error rates, coherence times, and gate speeds based on Tab.~\ref{tab:technologies}.}}
    \label{fig:technologies_example}
\end{wrapfigure}

To conclude hardware analysis, we investigate how different quantum technologies support the use of QEC codes on mid-term devices. Since each technology realizes qubits in a distinct way, this leads to heterogeneous noise models, gate speeds, and coherence times (\S~\ref{sec:background}), as shown in Fig.~\ref{fig:technologies_example}.

\myparagraph{Research question and hypothesis}
This leads us to ask \textbf{RQ\#5:} \emph{Which quantum technologies are most suitable for effective error correction, given their error rates and hardware restrictions?} We hypothesize that QEC codes are most effective on platforms that support shuttling, since enhanced connectivity plays a decisive role in error correction performance as we show in \S~\ref {subsec:connectivity}. Among these, trapped-ion devices appear particularly promising due to their long coherence times. However, noise rates in realistic mid-term hardware remain above most QEC thresholds, limiting the potential for full fault tolerance. Trapped-ion and neutral atom QPUs support mid-circuit movement and offer a higher ratio of gate speed to coherence time, while superconducting and neutral atoms devices show similar two-qubit error rates, with trapped-ion systems providing an order-of-magnitude improvement. Thus, we expect the mid-term noise levels to be too high for effective QEC.

\input{plots/eval_technologies/technologies}

\myparagraph{Methodology}  
We simulate QEC codes at the maximum distance on mid-term devices for each technology: Google Willow, Quantinuum Apollo, and Infleqtion. We use noise models that capture their corresponding error rates. For devices supporting shuttling, we compare performance with and without it.

\myparagraph{Results}
Fig.~\ref{fig:technologies} shows the logical error rate results. Quantinuum Apollo performs best, with average rates of $0.833 \times 10^{-3}$ without shuttling and $5.67 \times 10^{-3}$ with, followed by Infleqtion at $4.97 \times 10^{-1}$ without shuttling and $2.45 \times 10^{-1}$ with. The gross code fails in all cases except Apollo, where high connectivity and low error rates enable full error suppression. Across technologies, the concatenated Steane code performs best on average ($1.11 \times 10^{-1}$), although we still observed some errors on Apollo, likely due to the chosen concatenation level. For Apollo, only the concatenated Steane code degrades with shuttling, while all other codes maintain complete reduction. On average, shuttling improves performance from about 0.249 to 0.125, nearly halving the error rate. Projected trapped-ion error rates appear sufficient for practical QEC, as most codes eliminate logical errors on Apollo. On Infleqtion with shuttling, the surface code reaches $1.0 \times 10^{-3}$. Although current superconducting devices perform poorly, Fig.~\ref{fig:dqc} shows that a tenfold reduction in IBM Flamingo’s physical error rate enables its surface code to reach $9.0 \times 10^{-3}$. Our results indicate that if industry roadmaps \cite{IBM_roadmap, Quantinuum2024Roadmap} are met, fault tolerance may be achievable within five years.

\takeaway{\textbf{Takeaway \#5:} We observe that shuttling can nearly halve logical error rates. Among the platforms we study, Quantinuum Apollo (trapped-ion) performs the best, often presenting no errors in the sample, while on Infleqtion with shuttling, the surface code brings the logical error rate down to $1.0 \times 10^{-3}$. Our results hint that future low-error, high-connectivity hardware will enable near-complete error suppression, putting fault-tolerant QEC within reach.}

%% file: tables/quantum_error_codes_comparison.tex
\begin{table}[t]
\centering
\caption{Ratio of logical to physical qubits and number of two-qubit gates based on our implementations of the QEC codes. {\em Distance 11 for the majority of codes, 12 for the gross code, and 9 for the concatenated Steane with 3 syndrome measurement rounds.}}
\fontsize{8}{9}\selectfont
{
\begin{tabular}{|c|c|c|c|c|c|c|}
\hline
 & Surface & Bacon-Shor  & BB (Gross) & C. Steane & Color Code & Heavy-Hex \\
\hline
Net Enc. Rate & $\frac{1}{2d^2 - 1}$ & $\frac{1}{d^2}$ & $\frac{k}{2n}$ ($\frac{1}{24}$) \cite{Bravyi2024}& $\frac{1}{2 \cdot 7^m}$ & $\frac{4}{(3d-1)^2}$ \cite{Chamberland_2020_triangle} & $\frac{2}{(5d^2-2d-1)}$ \cite{Chamberland_2020} \\
\hline
\#Phys. Qubits & 274 & 121 & 288 & 98 & 181 & 291\\
\hline
\#2Q & 1320 & 1320 & 2592 & 961 & 1440 & 1920\\
\hline
\end{tabular}
}
\label{tab:code_stats}
\end{table}

%% file: plots/eval_connectivity/connectivity.tex
\begin{figure}[t]
    \centering
    \includegraphics[scale=0.44]{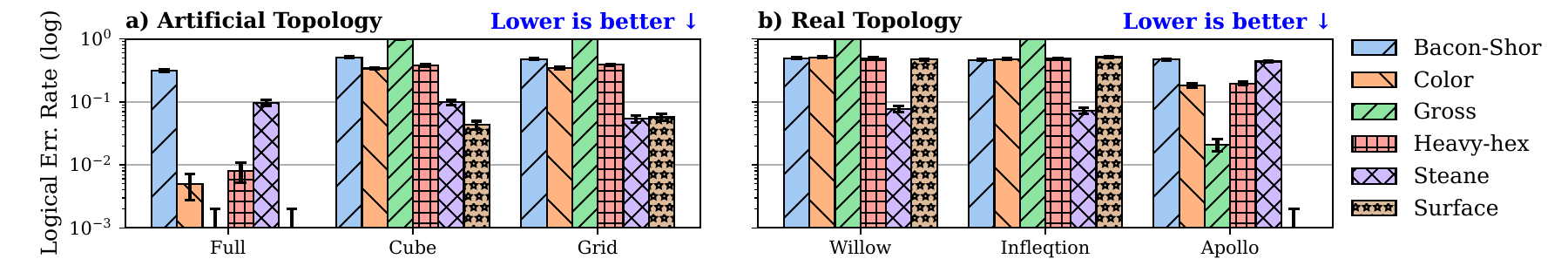}
    \caption{Logical error rates of different QEC codes under the SI1000 noise model across topologies with varying connectivity. {\em a) Artificial topologies of 300 qubits with different connectivities and error probability 0.002. b) Realistic topologies with error probability 0.004.}}
    \label{fig:connectivity}
\end{figure}

%% file: plots/eval_variance/variance.tex
\begin{figure}[t]
    \centering
    \includegraphics[scale=0.45]{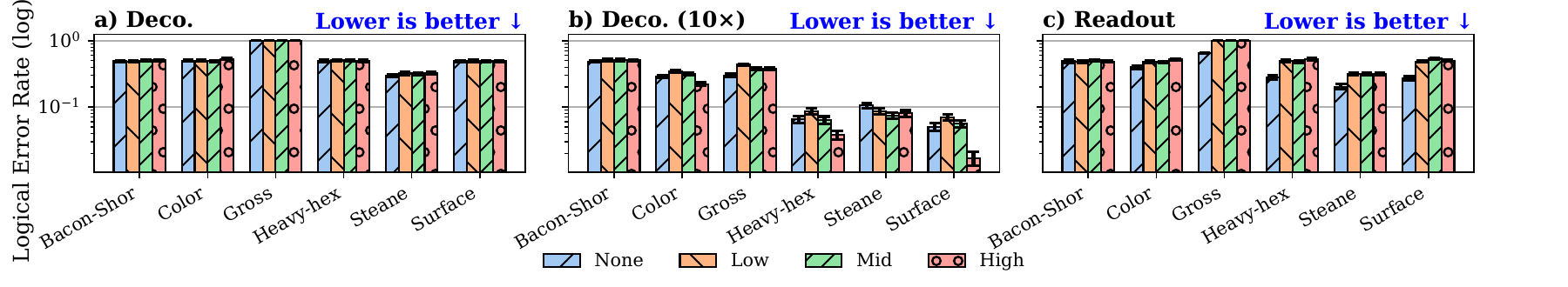}
    \caption{Influence of qubit quality and readout error variability on the QEC codes. {\em Comparison of logical error rates on an all-to-all topology for: (a) realistic qubit $T_1$ and $T_2$ sampled from normal distributions based on IBM Heron parameters at various variance levels, (b) tenfold prolonged $T_1$ and $T_2$, and (c) measurement errors with varying variance from IBM Heron.}}
    \label{fig:variance}
\end{figure}

%% file: plots/eval_dqc/dqc.tex
\begin{figure}[t]
    \centering
    \includegraphics[scale=0.45]{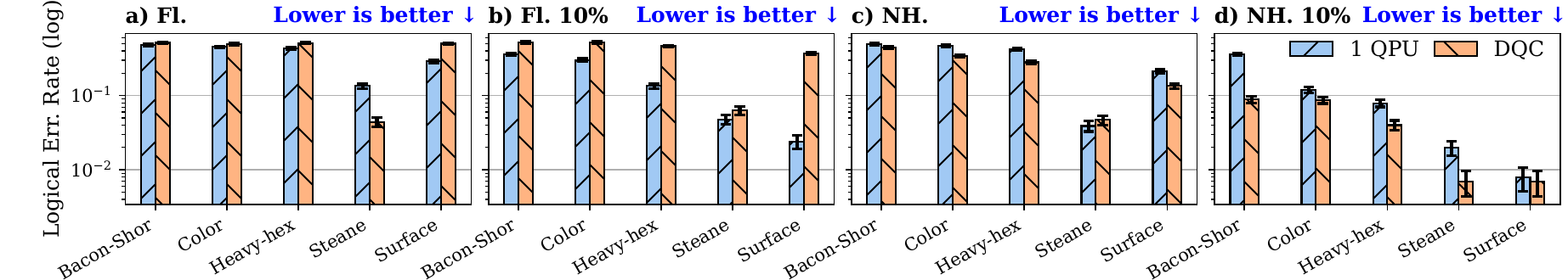}
    \caption{Performance of QEC codes on a single QPU vs. on the entire DQC architecture of IBM Flamingo (Fl.) and IBM Nighthawk (NH.) with original and tenfold downscaled noise.}
    \label{fig:dqc}
\end{figure}

%% file: plots/eval_technologies/technologies.tex
\begin{wrapfigure}{r}{0.6\textwidth}
    \centering
    \includegraphics[scale=0.44]{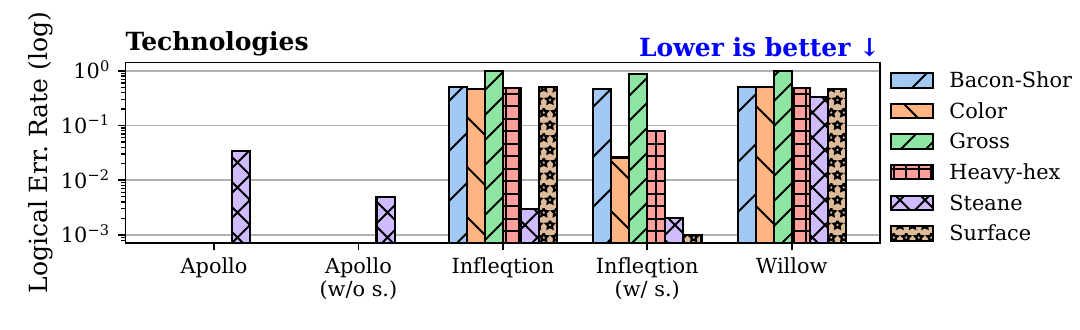}
    \caption{Effectiveness of codes on realistic mid-term devices, both w/ and w/o mid-circuit qubit movement.}
    \label{fig:technologies}
\end{wrapfigure}

%% file: chapters/framework_analysis.tex
\section{Framework Analysis}
\label{sec:compilation}
In this section, we analyze how certain compilation stages, such as mapping and routing (\S~\ref{subsec:mapping}) and translation to the target gate set (\S~\ref{subsec:translation}), affect the effectiveness of the protection offered by the chosen QEC codes. Here, we aim to provide insights into the error overhead that must be accounted for when selecting a QEC code for a QPU with a specific noise model and available compiler.

\subsection{Mapping and Routing}
\label{subsec:mapping}

\begin{wrapfigure}{r}{0.35\textwidth}
    \centering
    \includegraphics[scale=0.27]{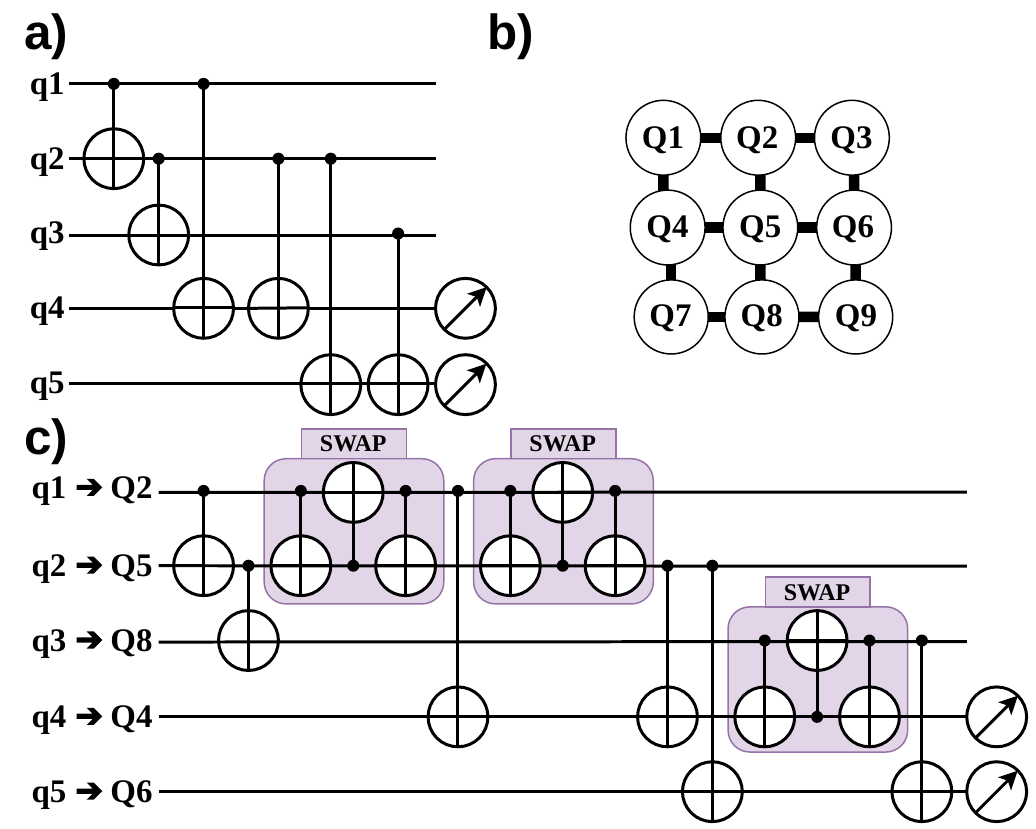}
    \caption{Circuit mapping and routing on an exemplary topology. {\em (a) Repetition code (d=3) (b) Grid topology (c) Exemplary mapping of logical to physical qubit, which necessitates additional SWAP gates.}}
    \label{fig:mapping_example}
\end{wrapfigure}

We investigate the impact of mapping and routing on QEC codes' performance, since these compilation stages substantially increase the number of noisy two-qubit gates \cite{maronese2021quantumcompiling}. Fig.~\ref{fig:mapping_example} shows this, where we map a circuit on a grid topology (a-b). The routing step that follows the mapping adds SWAP gates to resolve the connectivity constraints of the topology (c).

\myparagraph{Research question and hypothesis}
This motivates the research question:  
\textbf{RQ\#6:} {\em How do the mapping and routing compilation stages influence the effectiveness of QEC codes?} We hypothesize that when mapping and routing heuristics limit SWAP overheads, QEC effectiveness increases. Since many quantum devices do not support SWAPs natively \cite{AbuGhanem_2025, quantinuum_h2_datasheet_2024}, a SWAP is decomposed into three two-qubit gates, effectively tripling the error probability (Fig.~\ref{fig:mapping_example}(c)). 

\myparagraph{Methodology}  
We compile each code 1000 times  using all combinations of initial layout strategies and routing heuristics onto a $17 \times 17$ grid and the 193-qubit heavy-hex topology, where the gross code is excluded due to prohibitive runtime.

\begin{figure}[t]
    \centering
    \includegraphics[scale=0.42]{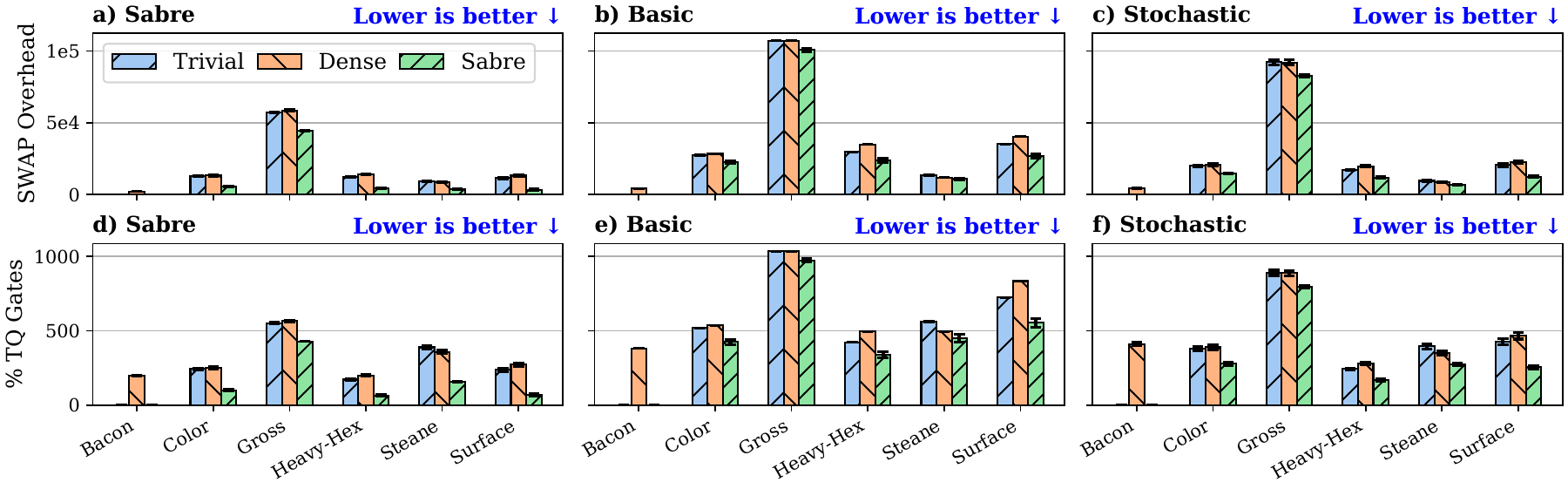}
    \caption{SWAP overhead (a-c) and SWAP overhead as percentage of original two-qubit gates (d-f) introduced for different QEC codes and routing methods on grid architecture.}
    \label{fig:mapping_grid}
\end{figure}

\begin{figure}[t]
    \centering
    \includegraphics[scale=0.42]{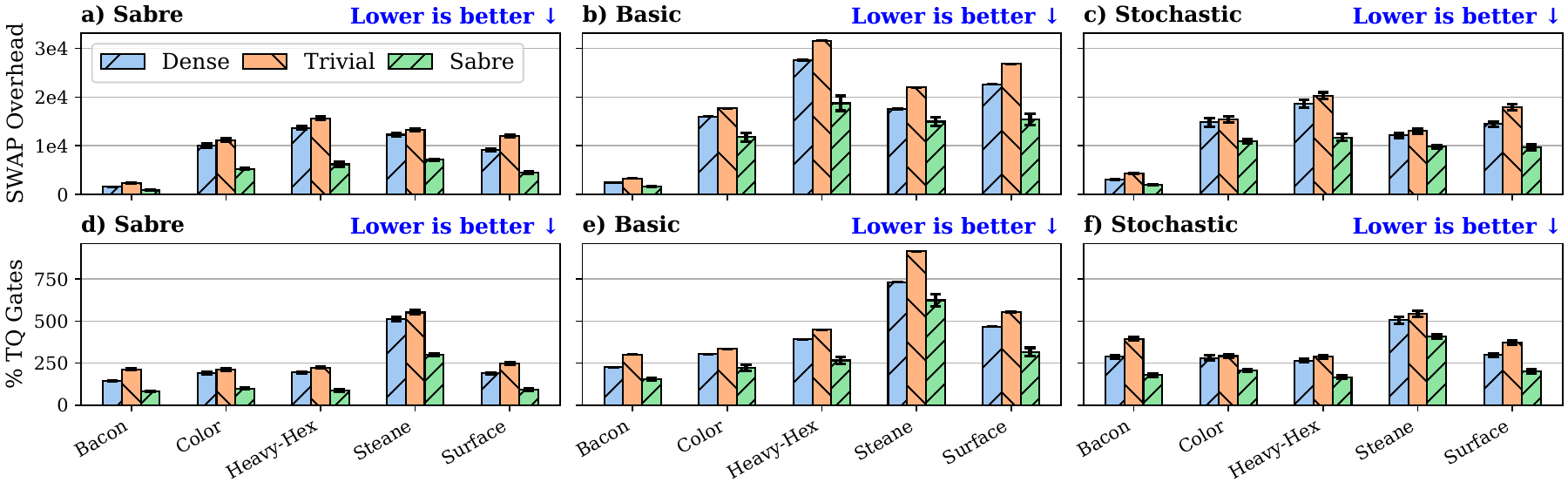}
    \caption{SWAP overhead (a-c) and SWAP overhead as percentage of original two-qubit gates (d-f) introduced for different QEC codes and routing methods on heavy-hex architecture.}
    \label{fig:mapping_hh}
\end{figure}

\myparagraph{Results}  
Figures~\ref{fig:mapping_grid}(a–c) and \ref{fig:mapping_hh}(a–c) show the SWAP overheads. We exclude the \texttt{Lookahead} mapping from our experiments due to its prohibitively long execution times on larger circuits and topologies. Across all codes, we find that SABRE yields the best results. On average, it introduces 10216.480 SWAP gates per code for grid topology, ranging from 0 for the Bacon–Shor code to 44415.49 for the gross code, and 4789.2756 SWAP gates per code for heavy-hex topology, ranging from 892.139 for the Bacon–Shor code to 7147.208 for the concatenated Steane code. However, even codes typically well-suited for the given topologies, such as the surface code for the grid or the heavy-hex code for the heavy-hex topology, accumulate substantial overhead (3288.283 and 6201.596 additional SWAPs, respectively). Notably, the overhead does not seem consistently proportional across all the codes. 

\myparagraph{Analysis} Next, we examine how SWAP overhead relates to the number of two-qubit gates in the original circuit, which we show in Figures \ref{fig:mapping_grid}(d–f) and \ref{fig:mapping_hh}(d–f). For the grid topology, SABRE incurs an average overhead of 136.34\%, reaching up to 428.39\% for the gross code. For the heavy-hex topology, SWAP overhead is around 131.85\%. Lastly, we observe that even for codes with similar connectivity needs, the proportions differ significantly. Because of the SWAP decomposition, in fact, we observe an average increase equivalent to 409.02\% and 395.55\% of the original two-qubit gates, respectively.

\takeaway{\textbf{Takeaway \#6:} We show that mapping and routing add 134.095\% more two-qubit gates due to SWAP overhead, on average. In practical applications, this corresponds to just over four extra two-qubit gates for every original one, which greatly amplifies two-qubit errors and highlights the need for mapping and routing strategies tailored to QEC to preserve fault tolerance.}

\subsection{Gateset Translation}
\label{subsec:translation}

\begin{wrapfigure}{r}{0.48\textwidth}
    \centering
    \includegraphics[scale=0.39]{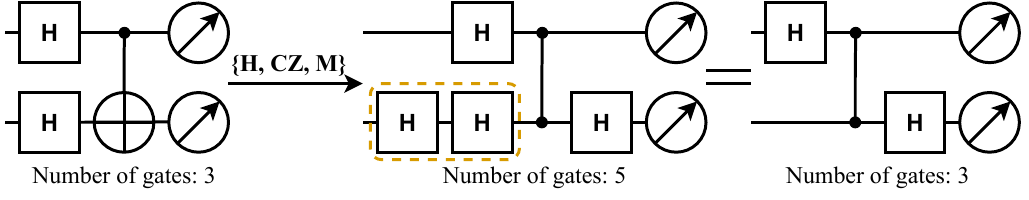}
    \caption{Example of circuit translation. {\em Non-optimal and optimal translation into a gate set w/o a CNOT gate.}}
    \label{fig:translating_example}
\end{wrapfigure}

We next analyze the error overhead introduced when translating a quantum circuit into a device's native gate set. Each quantum device supports a specific, restrictive set of hardware-level gates \cite{quantinuum_h2_datasheet_2024, AbuGhanem_2025}, requiring compilers to decompose high-level operations into hardware-compliant sequences. Fig.~\ref{fig:translating_example} illustrates this process, showing the decomposition required to implement a CNOT gate on a device that does not support it \cite{AbuGhanem_2025}.

\myparagraph{Research question and hypothesis} This raises a question: \textbf{RQ\#7:} {\em How does the translation stage influence the effectiveness of QEC codes?} We hypothesize that different compilers generate hardware-level gate sequences with varying overheads.
To illustrate our hypothesis, we consider the example in Fig.~\ref{fig:translating_example}. The same high-level circuit can be translated in multiple equivalent ways, including with or without optimization. These choices yield different numbers of gates, each contributing to the total error probability. 

\myparagraph{Methodology}  
We compare the number of additional gates introduced when translating each QEC code with maximal distance to the gate sets of IBM Heron and Quantinuum H2 devices. To  remove any connectivity constraints, we use a fully connected topology of 300 qubits.

\input{plots/eval_translating/gate_overhead}

\myparagraph{Results}  
We exclude BQSKit from our results since it removes operations deemed redundant, which interferes with error detection. In Fig.~\ref{fig:translating}(a-b), we show the number of additional gates introduced by translation. 
The gross code exhibits the highest absolute increase due to its large number of gates. On average, Qiskit introduces 27552 additional gates, while TKET adds over twice as many. At their highest optimization levels, Qiskit reduces the overhead by 21.94\%, while TKET at its highest viable level (L2 \cite{pytket_quantinuum_docs}) achieves just 1.7\%. These results demonstrate that optimization mitigates, but cannot fully eliminate, the cost of translation.

\myparagraph{Analysis} 
We investigate whether the overhead is not always proportional to the original gate count. In Fig.~\ref{fig:translating}(c-d), we show that, in contrast to routing, translation overhead scales approximately proportionally with the original gate count. Across all compilers and codes, an average of 6.134 additional gates is introduced per original gate, with TKET contributing 8.704, optimized TKET 8.406, Qiskit 4.261, and optimized Qiskit 3.166. We observe the highest deviation from the average in the Bacon-Shor code, with a normalized overhead of 5.301. This suggests that the majority of codes have similar distributions of gate types, leading to consistent translation scaling.

\takeaway{\textbf{Takeaway \#7:} Our investigation shows that even the most optimized translation introduces significant overhead, with an average of 3.166 additional two-qubit gates per original gate. This highlights that, compared to unoptimized circuits, careful compilation is crucial to minimize overhead and limit error accumulation, thereby preserving the effectiveness of QEC codes.}

%% file: plots/eval_translating/gate_overhead.tex
\begin{figure}[t]
    \centering
    \includegraphics[scale=0.45]{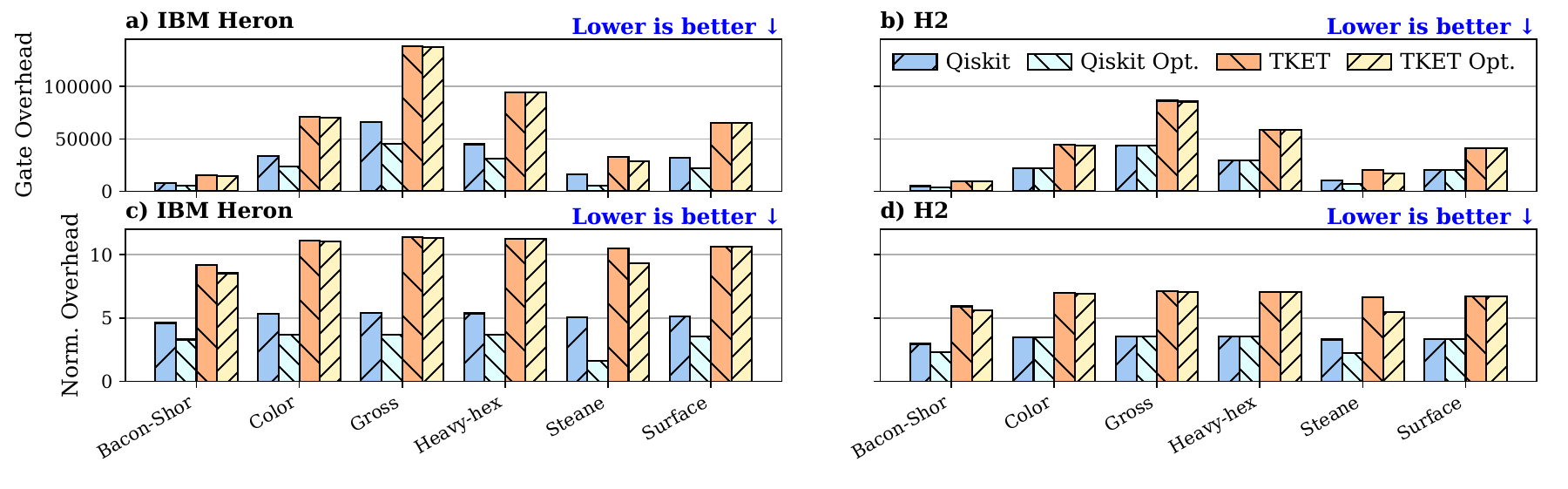}
    \caption{Gate overhead of converting to IBM Heron and Quantinuum H2 gate sets using TKET and Qiskit.}
    \label{fig:translating}
\end{figure}

%% file: chapters/noise_analysis.tex
\section{Quantum Error Correction Analysis}
\label{sec:qec_analysis}
Finally, we explore the correcting abilities of decoders (\S~\ref{sec:decoders}) and QEC codes (\S~\ref{sec:noise}) themselves.

\input{appendix_sections/decoders}
\subsection{Is QEC Essential for Quantum Computing?}
\label{sec:noise}

\begin{wrapfigure}{r}{0.5\textwidth} 
    \centering 
    \includegraphics[scale=0.33]{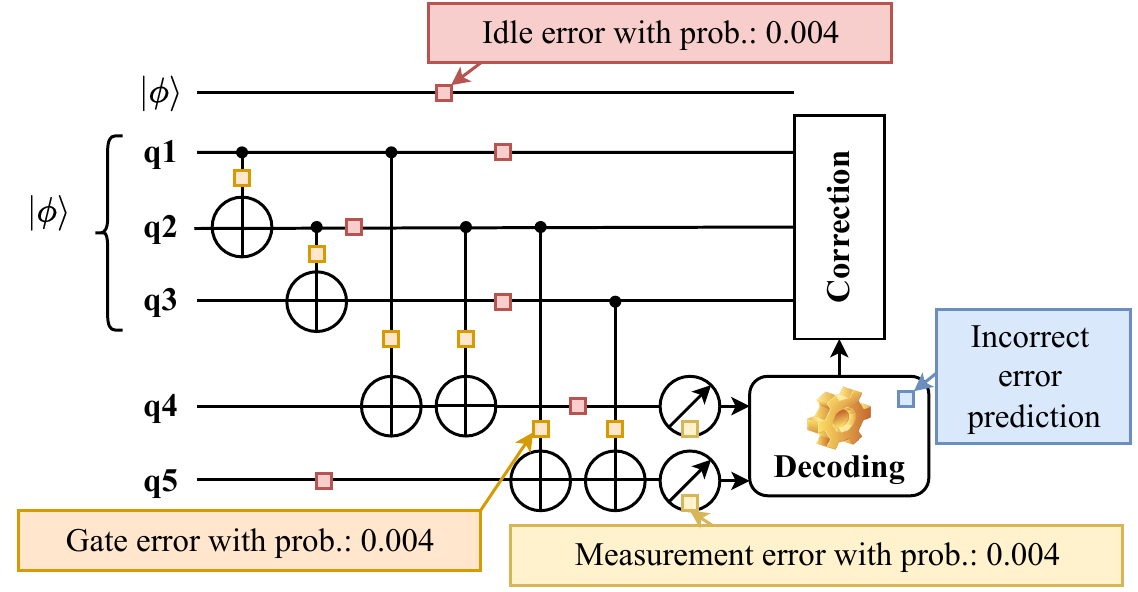} 
    \caption{Example of errors in an unprotected versus a protected qubit. 
    {\em Comparison of errors in quantum memory w/o protection and w/ a repetition code under an uniform noise model.}} 
    \label{fig:acc_example} 
\end{wrapfigure}

We close with a provocative question: is QEC always necessary? Here, we examine errors introduced by the correction procedure itself, highlighting the gap between ideal fault-tolerance and the reality of QPUs suffering from various noise sources (Fig.~\ref{fig:acc_example}).

\myparagraph{Research question and hypothesis}
Thus, we ask: \textbf{RQ\#9:} \emph{Is the application of QEC always beneficial, or are there regimes in which the overhead of QEC introduces more noise than it suppresses?} We hypothesize that applying QEC can be detrimental as the error probability approaches a code’s threshold. Consider a single idle qubit with error probability $p_{idle}$. Protecting it with a simple repetition code (Fig.~\ref{fig:acc_example}) introduces additional errors from gates ($p_{gate}$) and measurements ($p_{meas}$), yielding a combined error probability of approximately $5 \times p_{idle} + 6 \times p_{gate} + 2 \times p_{meas}$. Once the code’s correcting capabilities are exceeded, the correction process itself becomes a net source of faults, making the encoded qubit noisier than the unprotected one. We aim to investigate whether, across different error probabilities and especially near the threshold, this extra overhead causes the logical error rate to grow faster than that of the unprotected qubit.

\myparagraph{Methodology} We compare error growth in protected and unprotected circuits by generating the QEC codes at maximal distance, on a fully connected 300-qubit topology, and their corresponding unprotected circuits, consisting of a single idle qubit (12 for the gross code), which we keep idle for as many rounds as their protected counterparts. We add noise using the SI1000 model with two-qubit error probabilities in the 0.001–0.01 range, and record the resulting error rates.

\input{plots/thresholds/thresholds}

\myparagraph{Results} We observe no errors on unprotected qubits at any probability. We present results for protected circuits in Fig.~\ref{fig:thresholds}. At $p=0.001$, we observe errors for two codes. At $p=0.004$, all codes exceed their thresholds. Surface code exhibits the slowest error growth, while the gross code, with the most gates, deteriorates most sharply. We expect that in more constrained topologies, the higher gate count will result in even faster performance degradation.

\myparagraph{Analysis} To further investigate the introduced overhead, we also isolate the effect of the correction itself. As expected, logical error rates are almost always higher without correction, even after a code's threshold is crossed, confirming the codes' error-suppressing capabilities. However, as shown in Tab.~\ref{tab:higher_raw_error}, this benefit is not universal: in some cases, incorrect decoder decisions actively degrade performance rather than improve it.

\input{tables/higher_raw_error}

\takeaway{\textbf{Takeaway \#9:} Our study shows that QEC is not universally beneficial. At $p=0.001$, we observe no errors only on four of the evaluated codes, and at $p=0.004$ all codes cross their thresholds, while unprotected qubits still show no observed errors. Although QEC can suppress errors at low rates, once a code’s threshold is crossed, its performance deteriorates rapidly, depending on its correcting capabilities and the noise sources in the circuit. These results suggest that QEC should be applied selectively, only where error rates are low or to protect complex circuits rather than quantum memory, to achieve meaningful error suppression.}

%% file: appendix_sections/decoders.tex
\subsection{Decoders}
\label{sec:decoders}

\begin{wrapfigure}{r}{0.4\textwidth}
    \centering
    \includegraphics[scale=0.3]{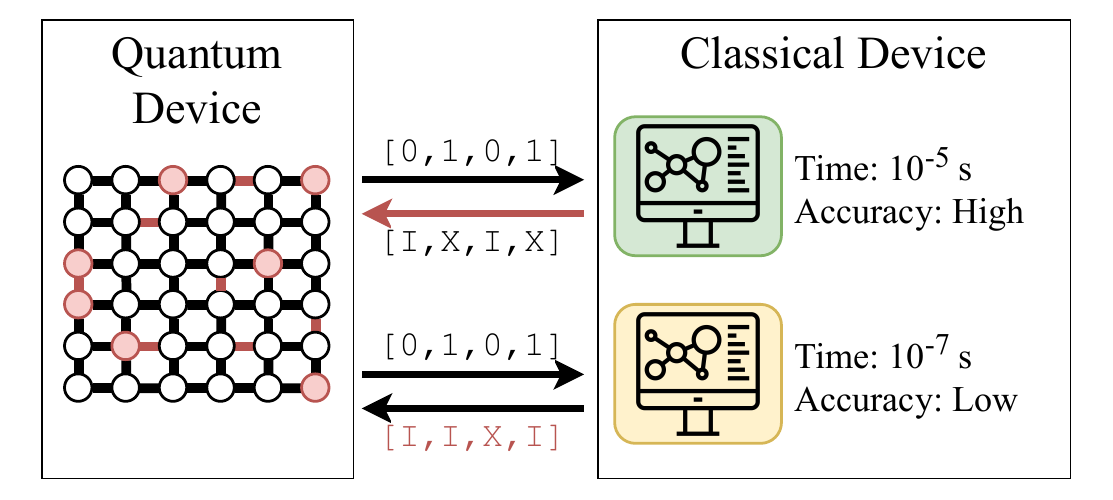}
    \caption{Decoding algorithms demonstrate a trade-off between speed and accuracy.}
    \label{fig:decoders_example}
\end{wrapfigure}

We examine how the choice of decoder shapes QEC performance, as these algorithms interpret syndrome measurements, predict errors, and propose corrections, with varying accuracy and efficiency \cite{maan2023testingaccuracysurfacecode, Battistel_2023} (Fig.~\ref{fig:decoders_example}).

\myparagraph{Research question and hypothesis}
This leads us to raise a question \textbf{RQ\#8}: {\em Which decoders achieve the best performance across QEC code families?} We hypothesize that no single decoder consistently dominates, as performance depends on code structure and error propagation.

\begin{figure}[t]
    \centering
    \includegraphics[scale=0.4]{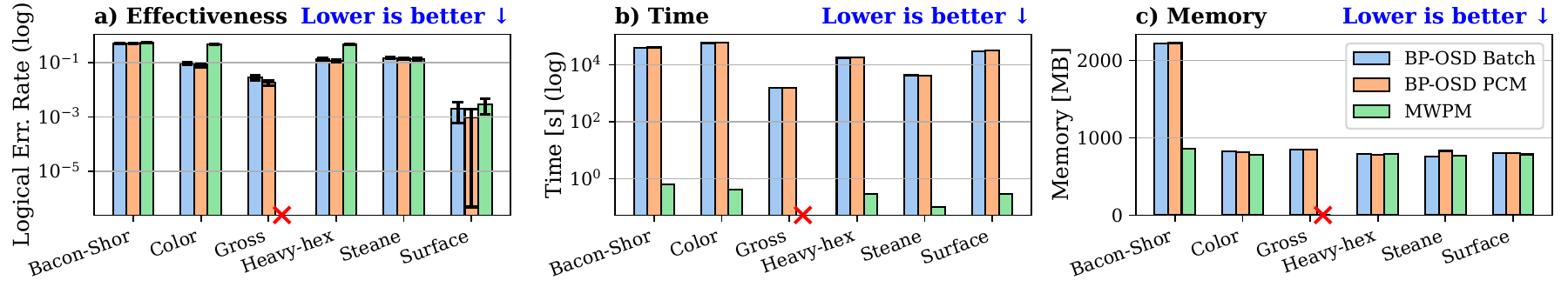}
    \caption{Comparison of decoders. {\em a) Effectiveness against the SI1000 noise model with probability 0.004. b) Runtime of the decoders. c) Memory usage of the decoders.}}
    \label{fig:general_decoders}
\end{figure}

\begin{figure}[t]
   \centering
    \includegraphics[scale=0.4]{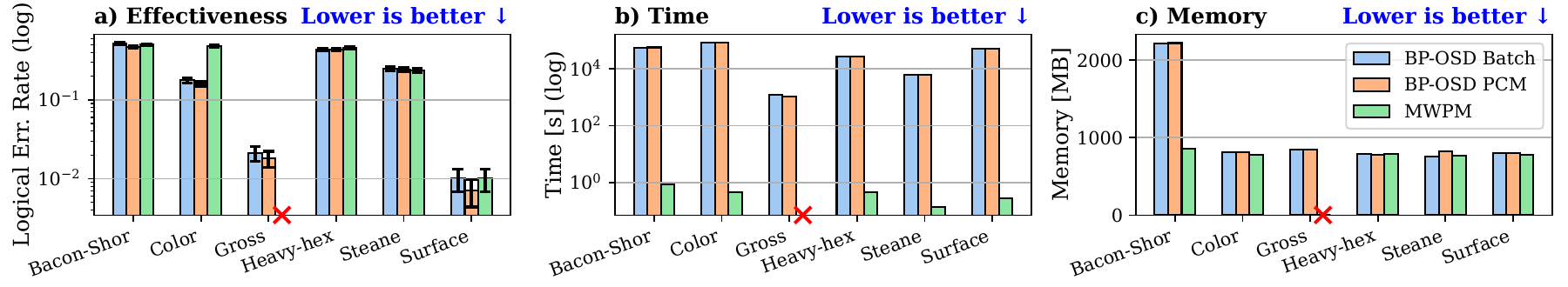}
    \caption{Comparison of decoders. {\em a) Effectiveness of different decoders against the uniform noise model with probability 0.004. b) Runtime of the decoders. c) Memory usage of the decoders.}}
    \label{fig:general_decoders_const}
\end{figure}

\myparagraph{Methodology}
We evaluate two open-source decoders: \textit{BP-OSD} \cite{bposd_roffe_2020} and \textit{MWPM} \cite{pymatching_higgott_2025}, across multiple QEC codes under the SI1000 and the uniform error models ($p=0.004$). BP-OSD is tested in two variants: parity-check matrix decoding and batched error decoding. In both cases, we use minimum-sum belief propagation with up to 10000 iterations for improved convergence, followed by order-7 OSD, trading increased accuracy for higher computational cost. Each configuration runs on a fully connected 400-qubit topology, using the maximum achievable code distance for each QEC scheme.

\myparagraph{Results}
Excluding the gross code, for which MWPM fails, the BP-OSD parity-check decoder achieves the best overall performance against the SI1000 noise model across the tested QEC codes, with an average logical error rate of 0.168, outperforming MWPM by 46.66\% on average (Fig.~\ref{fig:general_decoders}). MWPM outperforms BP-OSD only for the concatenated Steane code. However, this improved accuracy comes at a substantial computational cost: MWPM requires, on average, only 0.34~s per experiment, whereas BP-OSD requires over 8~h, and MWPM uses 26.89\% less memory than the heavier parity-check BP-OSD implementation.   Fig.~\ref{fig:general_decoders_const} presents the results for the uniform noise model. Here, on average, the BP-OSD decoder with a parity-check matrix achieves the best overall performance across all tested QEC codes, yielding an average logical error rate of 0.2624. It outperforms BP-OSD batch (0.2764) and MWPM (0.3378). Again, the average execution time for BP-OSD algorithms is over $10^5$ times longer, while the memory usage is 26.86\% smaller for the MWPM than for the less resource-efficient BP-OSD with a parity-check matrix. These results reflect the choice of aggressive BP-OSD settings optimized for decoding accuracy. Reducing the number of BP iterations or the OSD order would significantly shorten the runtime at the expense of performance.

\begin{figure}[t]
    \centering
    \includegraphics[scale=0.41]{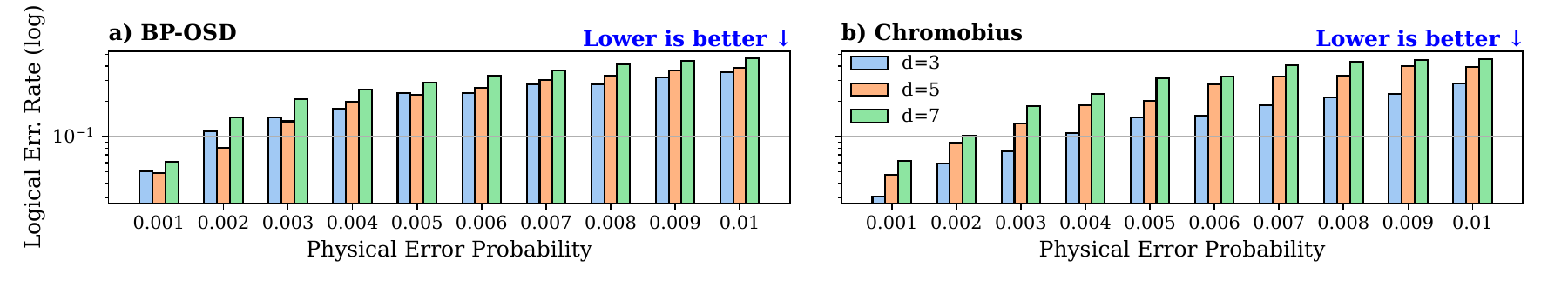}
    \caption{Difference in logical error rate between BP-OSD and Chromobius for color codes of various distances under the SI1000 error model.}
 \label{fig:color_decoders}
\end{figure}

\begin{figure}[t]
    \centering
    \includegraphics[scale=0.41]{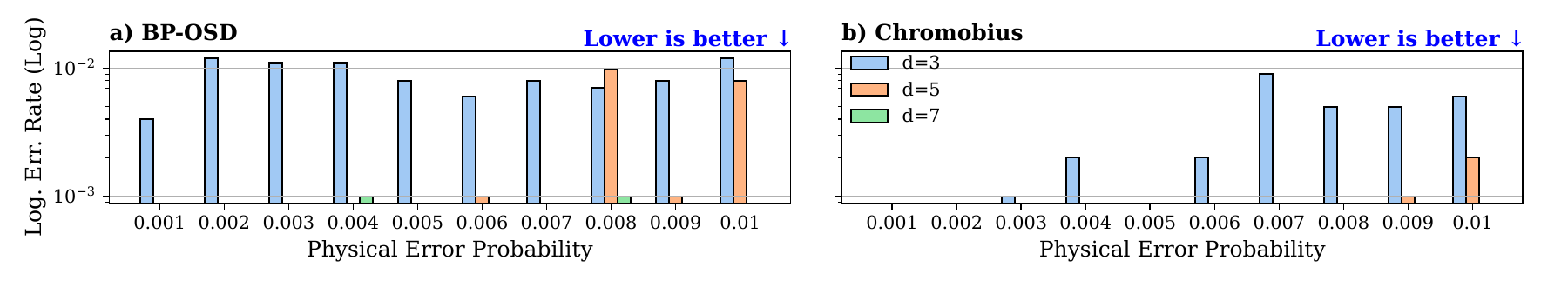}
    \caption{Difference in logical error rate between BP-OSD and Chromobius for color codes of various distances under the phenomenological error model.}
    \label{fig:color_decoders_ph}
\end{figure}

\myparagraph{Analysis}
General-purpose decoders are essential for cross-code benchmarking but may underperform compared to specialized decoders. We compare the performance of the BP-OSD batch and \textit{Chromobius} \cite{gidney2023newcircuitsopensourceChromobius} decoder tailored for the color code, at distances 3, 5, and 7. Under the SI1000 noise model, Chromobius achieves an average logical error of 0.228 versus BP-OSD’s 0.289 (Fig.~\ref{fig:color_decoders}), showing an advantage, but smaller than we anticipated. We further compare the performance of the color code using Chromobius and BP-OSD under a standard, albeit less realistic, phenomenological noise regime, where only noise sources are the depolarization of data qubits between rounds and imperfect measurements \cite{gidney2023newcircuitsopensourceChromobius}. In Fig.~\ref{fig:color_decoders_ph}, we show that under this noise model, Chromobius achieves an average logical error rate of 0.0011, performing approximately 3.3$\times$ better than BP-OSD.

\takeaway{\textbf{Takeaway \#8:} BP-OSD achieves the best general decoding performance and, on color code under a realistic noise model, performs only 26.75\% worse than a dedicated decoder, making it the most suitable choice for cross-code evaluations.}

%% file: plots/thresholds/thresholds.tex
\begin{wrapfigure}{r}{0.4\textwidth}
    \centering
    \includegraphics[scale=0.45]{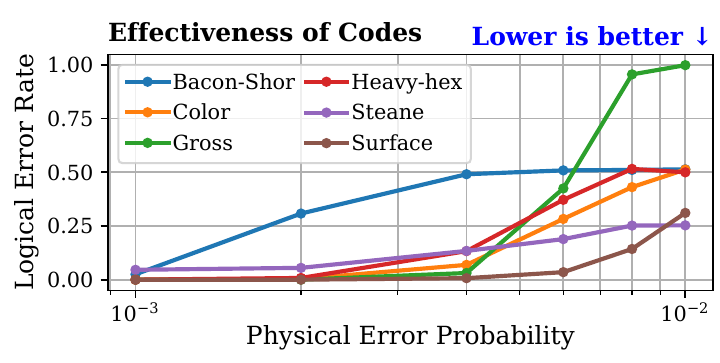}
    \caption{Logical error rates for different codes under the SI1000 noise model of various probabilities in a realistic setup.}
    \label{fig:thresholds}
\end{wrapfigure}

%% file: tables/higher_raw_error.tex
\begin{table}[t]
\centering
\caption{Observed increase in logical error rate due to decoding.}
\fontsize{8}{9}\selectfont
\begin{tabular}{|l|cccc|cc|cc|}
\hline
QEC Code
 & \multicolumn{4}{c|}{Bacon--Shor}
 & \multicolumn{2}{c|}{Heavy--Hex}
 & \multicolumn{2}{c|}{Con.~Steane} \\
\hline
Error prob.
 & 0.004 & 0.006 & 0.008 & 0.010
 & 0.008 & 0.010
 & 0.001 & 0.008 \\
\hline
$\Delta \mathrm{LER} = \mathrm{Corrected} - \mathrm{Uncorrected}$
 & +0.003 & +0.002 & +0.004 & +0.005
 & +0.016 & +0.010
 & +0.014 & +0.012 \\
\hline
\end{tabular}
\label{tab:higher_raw_error}
\end{table}

%% file: chapters/limitations.tex
\section{Limitations}
\label{sec:limitations}

\myparagraph{\projectname{} implementation}
Our implementation does not include FT gate constructions, as no general, scalable framework exists across different QEC codes, and practical methods to enable FT gates are still an active area of research aimed at reducing overhead. Once such methods become available, \projectname{} will support circuits in this more general FT setting. We also omit realistic decoding delays, since latency depends strongly on the QEC code, decoder, and hardware, making general comparisons difficult. Although we could assume optimistic sub-microsecond latencies, they would have minimal impact and risk introducing code- or hardware-specific bias. Future versions of \projectname{} may offer configurable latency models.

\myparagraph{\projectname{} execution}
\projectname{} supports exact sampling of QEC-protected quantum memory circuits, but at significant memory and runtime cost. Fig.~\ref{fig:program_stats} shows that the decoding stage, during which errors are sampled via stabilizer simulation and a decoding heuristic is applied, dominates runtime, accounting for an average of 99.96\% of runtime, as errors are sampled via stabilizer simulation and processed by a decoding heuristic. Since Stim can simulate a surface code ($d=100$) in 15~s \cite{Gidney_2021}, decoding becomes the bottleneck as detection events increase with code distance and additional gates for sparse backends, with shot count determining total runtime. Memory usage remains consistent across execution stages and scales with backend size, connectivity, and shots, reaching up to 815~MB per experiment. Overall, \projectname{} execution is constrained by both memory and runtime, which can extend over multiple days.

\input{plots/program_stats/program_stats}

%% file: plots/program_stats/program_stats.tex
\begin{figure}[t]
    \centering
    \includegraphics[scale=0.4]{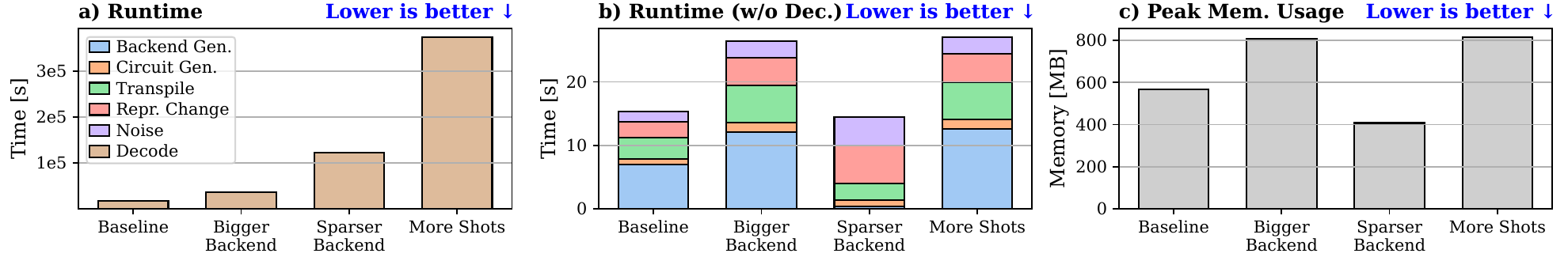}
    \caption{Runtime per stage and peak memory usage of \projectname{}.
    {\em Baseline: Running surface code on a fully connected 300-qubit backend (SI1000 noise, error 0.004) with BP-OSD decoder for 1000 shots. Bigger backend: 400 qubits. Sparser backend: grid topology. More shots: 10000 shots.}}
    \label{fig:program_stats}
\end{figure}

%% file: chapters/related_work.tex
\section{Related Work}

\myparagraph{Benchmarking in quantum computing}
Recent comprehensive surveys \cite{benchmarking_lorenz_2025, benchmarking_survey_proctor_2024} highlight the crucial role of benchmarking in guiding the development of the entire quantum computing stack. Established suites now exist for assessing quantum software \cite{benchmark_qasmbench_2022, Quetschlich_2023, benchmark_supermarq_2022}, development kits \cite{benchpress_2025}, and physical qubit performance \cite{ibm_layer_fidelity_2023}. However, these existing benchmarking studies are QEC-agnostic and thus orthogonal to our work.

\myparagraph{Benchmarking in QEC}
While systematic evaluations of QEC codes are scarce, existing studies tend to be limited in scope. Much of the prior work focuses on specific QEC-related aspects \cite{bechmarking_controller_kurman_2024, benchmarking_ml_qec_zhao_2024, surface_xzzx_vs_decoders_2023, resource_estimation_suchara_2013, steane_reed_vs_surface_2017, qecc_synth_2025}, relies on theoretical analysis with simplified noise models \cite{golay_vs_hamming_2003, comparison_11codes_2009, comparison_many_codes_huang_2019, steane_vs_surface_2017, comparison_honeycomb_vs_surface_2021, Bravyi2024, surface_vs_color_2007, comparison_surface_vs_color_2024}, or tests only a single code, topology, or noise model under realistic conditions \cite{surface_vs_realistic_noise_2014, qpandora_chatterjee_2025, steane_noise_comparison_2015, comparison_hypergraph_qldpc_2020, estimated_concantanated_2022, heavyhex_2025, bacon_surface_shor_trapped_ion_2020}. Furthermore, evaluations on real hardware remain rare \cite{small_codes_on_hardware_2022, comparison_shor_vs_steane_on_beacon_trappedion_2024}. In contrast, our work bridges this gap by systematically benchmarking a wide range of QEC codes under realistic, device-specific noise models, while accounting for compilation effects to deliver a comprehensive evaluation of their practical applicability.

\myparagraph{Existing frameworks}
While several QEC benchmarking frameworks exist, they are often limited. The most established is Stim \cite{framework_stim_2021}, which enables efficient stabilizer simulations but offers only simple noise models out of the box. Other efforts are either limited in scope \cite{framework_wille_2023}, deprecated \cite{framework_wootton_2020}, or restricted to specific code families \cite{framework_kang_2025, framework_mitsuki_2024}. In contrast, our work builds on Stim’s stabilizer simulation to provide a general framework for evaluating a broad spectrum of codes under precise noise models, while incorporating software stack effects and maintaining efficiency.

\myparagraph{Surveys}
While several surveys of QEC codes exist, they often focus on deep theoretical foundations \cite{survey_matsumoto_2021, survey_noise_adapted_2022, theoretical_overview_2023} or target broader audiences \cite{qec_for_dummies_2023, magic_mirror_2024, Roffe_2019}. Even the comprehensive Error Correction Zoo \cite{ErrorCorrectionZoo}, a valuable community catalog, uses a non-uniform presentation that hinders direct comparison. In contrast, our work focuses on the practical aspects of QEC and introduces a clear taxonomy designed to highlight the key characteristics and interrelations between different code families, enabling a more direct, comparative analysis.

\takeaway{\textbf{How our paper differs?}
This work systematically benchmarks multiple QEC codes with an emphasis on practical application, evaluating them under real devices, compilers, and noise models. To do so, we introduce a framework that integrates diverse codes, architectures, compilers, noise models, and decoders, and is easily extensible as the field evolves. Complementing this, our survey targets a broad audience and introduces a clear taxonomy that highlights code characteristics, overlaps, and differences, clarifying their strengths and constraints.
}

%% file: chapters/future_implications.tex
\section{Concluding Remarks}

We present a unified benchmarking framework enabling systematic and comprehensive empirical analysis of QEC codes. Our results provide valuable insights into the performance of QEC codes under various hardware- and framework-specific limitations, aiding the pursuit of FT quantum hardware. Future work should investigate factors beyond the scope of this study, including the evaluation of decoders and the translation of quantum circuit gates to their FT counterparts.

Having systematically explored the performance of QEC codes across mid-term devices, topologies, and noise regimes, we now draw practical lessons for hardware design, compilation strategies, and execution of quantum circuits. Our findings highlight which factors matter most for effective error correction and which can be deprioritized.

\begin{itemize}[leftmargin=*]
    \item \textbf{Trapped-ion in the lead:}
    Predicted trapped-ion devices with shuttling achieve the strongest error suppression, nearly eliminating logical errors. Prioritizing this platform's development is the most direct path to achieving fault-tolerant QEC within five years (\textbf{Takeaway \#5}).

    \item \textbf{Connectivity outweighs code size:}
    Enhancing qubit connectivity through features like mid-circuit movement or shuttling dramatically reduces logical errors, proving far more effective than simply increasing code distance (\textbf{Takeaways \#1, \#2, \#5}).
    
    \item \textbf{Variability is less critical:} Variations in qubit or readout quality have minimal impact on logical errors. This suggests that variance-aware mapping or code selection can be de-emphasized, simplifying both compilation and operational planning (\textbf{Takeaway \#3}).
    
    \item \textbf{Distributed execution is still challenging:} Single-QPU patches consistently outperform distributed setups due to the limited connectivity of inter-QPU couplers. Increasing the number of links is the key to enabling distributed fault-tolerant QEC (\textbf{Takeaway \#4}).
    
    \item \textbf{QEC-aware compilation is crucial:}
    The compilation process, particularly routing and mapping, can quadruple the two-qubit gate count. QEC-aware compilation is therefore essential to minimize this error-inducing overhead and preserve a code's effectiveness (\textbf{Takeaways \#6, \#7}).
    
    \item \textbf{QEC is not always helpful:} In our experiments, we found that in regimes where gate errors dominate, applying QEC indiscriminately can actually introduce more errors than it corrects. For near-term devices, this means we need to carefully decide when and where to apply QEC, particularly avoiding protection of idle qubits unless necessary, and focusing on more complex circuits where error correction truly provides a benefit (\textbf{Takeaway \#9}).

\end{itemize}

\myparagraph{Artifact} \projectname{}, along with the entire experimental setup, datasets, and results,
is publicly available on Zenodo:
\url{https://doi.org/10.5281/zenodo.19241564}.